\documentclass[11pt,english]{article}
\usepackage{titlesec}
\usepackage[T1]{fontenc}
\usepackage[latin1]{inputenc}
\usepackage{geometry}
\usepackage{amsmath}
\usepackage{setspace}
\usepackage{floatrow}
\usepackage{graphicx,color}
\usepackage{appendix}
\usepackage{multirow}
\usepackage{tikz}
\usepackage{babel}
\usepackage{enumitem}
\usepackage[hang,bf]{caption}
\usepackage{float}
\usepackage{amsthm}
\usepackage{footmisc}
\usepackage{amsmath}

\newtheorem{theorem}{Theorem}
\newtheorem{assumption}{Assumption}
\newcommand\numberthis{\addtocounter{equation}{1}\tag{\theequation}}

\begin{document}

\title{\vspace{-2.0em} Under-Identification of Structural Models Based on
Timing and Information Set Assumptions}
\author{{\normalsize Daniel Ackerberg} \\
{\normalsize Department of Economics }\\
{\normalsize University of Texas at Austin }\\
\and {\normalsize Garth Frazer} \\
{\normalsize Rotman School of Management}\\
{\normalsize University of Toronto}\\
\and {\normalsize Kyoo il Kim} \\
{\normalsize Department of Economics}\\
{\normalsize Michigan State University}\\
\and {\normalsize Yao Luo} \\
{\normalsize Department of Economics}\\
{\normalsize University of Toronto}\\
\and {\normalsize Yingjun Su\thanks{{\small Corresponding author: Yingjun
Su, Associate Professor, Institute for Economic and Social Research, Jinan
University, Guangzhou, China; Email: yis17@pitt.edu. \
We thank Victor Aguirregabiria, Manuel Arellano, Ariel Pakes and Peter Schmidt for helpful comments. All errors are our own.}}} \\
{\normalsize Institute for Economic and Social Research}\\
{\normalsize Jinan University}}
\date{{\normalsize June 2022 \vspace{-1em}}}

\maketitle
\thispagestyle{empty}

\begin{abstract}
\noindent We revisit identification based on timing and information set assumptions in structural models, which have been used in the context of production functions, demand equations, and hedonic pricing models (e.g. Olley and Pakes (1996), Blundell and Bond (2000)). First, we demonstrate a general under-identification problem using these assumptions in a simple version of the Blundell-Bond dynamic panel model. In particular, the basic moment conditions can yield multiple discrete solutions: one at the persistence parameter in the main equation and another at the persistence parameter governing the regressor. We then show that the problem can persist in a broader set of models but disappears in models under stronger timing assumptions. We then propose possible solutions in the simple setting by enforcing an assumed sign restriction and conclude by using lessons from our basic identification approach to propose more general practical advice for empirical researchers.
\end{abstract}

\singlespacing

\thispagestyle{empty}

\restoregeometry

\clearpage

\section{Introduction}

A number of recent papers address endogeneity in structural models using a
panel data identification strategy based on timing and information set
assumptions. \ A preeminent example is estimation of production functions
with endogenous input choice - in this context, timing and information set
assumptions are used for identification in two distinct lines of literature:
estimation methods using a proxy variable approach (e.g. Olley and Pakes
(1996), Levinsohn and Petrin (2003), and Ackerberg, Caves, and Frazer (ACF,
2015)), and estimation based on what are often called dynamic panel methods
(e.g. Blundell and Bond (2000)). \ The methods proposed in these literatures
have been used in thousands of papers, both in the context of production
functions and in many other structural models. In this paper, we study an
interesting identification problem that can arise in these methods. \ This
under-identification issue has been pointed out more narrowly in prior
papers (see footnote 16 of ACF and Kim, Luo, and Su (KLS, 2019)) which
concern the proxy variable approach. \ In this paper, we illustrate how this
is a more widespread issue in estimation methods that rely on timing and
information set assumptions. \ In particular, we show that the
under-identification problem can also arise in the context of dynamic panel
methods. \ Because dynamic panel methods have a relatively simple
representation, we then use this dynamic panel context to further study the
under-identification problem and suggest possible solutions.

The identification problem we study is unusual. \ While typical
identification problems involve regions of non-identification, this
identification problem does not - in our simple models, the moment
conditions are zero at the true parameters and non-zero locally around the
true parameters, but also at a second distinct point in the parameter space.
\ So, the model might be described as being locally identified (around the
true parameters) but not globally identified. \ There is a quadratic aspect
to the model and moment conditions that leads to the two possible solutions.
\ This form of the non-identification also impacts some of the solutions we
propose for the issue - in particular it means that we can leverage sign
restrictions to resolve it.

Before proceeding, we discuss a couple of important issues. \ First, we
illustrate the under-identification problem with perhaps the simplest
possible dynamic panel model. \ In particular, our basic model does not
include fixed effects, which are relatively common in the empirical
literature using these techniques. \ This is mainly to aid in our initial
illustration of the problem and solutions. \ Later in the paper, we show
that the under-identification problem also exists in dynamic panel models
with fixed effects.

Second, much of our discussion is couched in the production function
context. This is because production functions are perhaps the most frequent
empirical context for the application of these identification strategies
based on timing and information set assumptions. \ However, it is important
to note that there are many empirical applications of dynamic panel methods
(and timing/information set assumptions more generally) in other scenarios.
One interesting example is a set of recent work on demand estimation with
potential endogenous product characteristics, e.g. Sweeting (2009), Grennan
(2013), Lee (2013), and Sullivan (2017). Another example is some recent work
on hedonic price regression models with a time-varying unobserved attribute
that is correlated with observed attributes and follows a persistent
time-series process, e.g. Bajari, Fruehwirth, Kim, and Timmins (2012) and
Bishop and Murphy (2019). This under-identification issue and potential
solutions we propose are potentially relevant for all this work as well.

Our findings are also related to Sentana (2015), who also studies
\textquotedblleft finite\textquotedblright\ under-identification problems.
While he considers general methods to efficiently estimate all
observationally equivalent parameters that are consistent with an underlying
true model, we are instead interested in singling out the true parameter by
imposing an additional assumption in our specific model. \ Our point is that
enforcing this additional assumption selects the true parameter that is
consistent with the DGP, while the moment conditions typically used in
practice yield multiple solutions. Sentana (2015)'s examples of finite
under-identification are also distinct from ours, and an important
contribution of our paper is to illustrate the under-identification in this
particular model - a type of model that has been very widely used in the two
large literatures cited above.\footnote{%
One of the examples in Sentana (2015) is derived from the dynamic panel
literature (e.g. Arellano and Bond (1991)), but it concerns the model
\begin{equation*}
Y_{it}=\alpha _{1}\left( Y_{i,t-1}-\eta _{i}\right) +\alpha _{2}\left(
Y_{i,t-2}-\eta _{i}\right) +\eta _{i}+\upsilon _{it}
\end{equation*}%
and an under-identification issue involved in identifying $\alpha _{1}$ and $%
\alpha _{2}$, which arises when $\alpha _{1}+\alpha _{2}=1$. This is
distinct from the model we study, which is used in the empirical literature
to assess the impact of an (partially) endogenous $X_{it}$ on $Y_{it}.$ And
as noted in the text, our goal is to study how one can resolve the
identification problem (both in theory and in practice) by imposing
additional economic restrictions, unlike Sentana (2015) who studies the full
properties of the finite identified set.}

The paper proceeds by first illustrating the under-identification problem in
what is perhaps the simplest possible dynamic panel model. We then broaden
our scope to consider the problem more generally, showing how it can
manifest itself in more general models (e.g. dynamic panel models with fixed
effects, different data generating processes for the right-hand-side
endogenous variables, and models with multiple right-hand-side endogenous
variables). We then turn to potential solutions to the problem, in part
based on specifying a model for the endogenous right-hand-side variables
(e.g. inputs in the production function context) and enforcing an assumed
sign restriction. We end by using lessons from our specific solutions for
this simple example to propose more general advice for researchers using
these methods (both proxy variable and dynamic panel approaches) in practice
- in an effort to hopefully minimize the problems stemming from this
under-identification result.

\newpage
\section{\protect\bigskip The Under-Identification Problem}

\label{sec:id-problem}

Consider a panel regression model
\begin{equation}
y_{it}=\alpha ^{0}+\beta ^{0}x_{it}+\omega _{it}+\eta _{it},  \label{eq:pf0}
\end{equation}%
where $x_{it}$ is correlated with error term $\omega _{it}$ but not with $%
\eta _{it}$. Our leading example is production function estimation, in which
$y_{it}$ is the output, $x_{it}$ is an input, $\omega _{it}$ is a serially
correlated unobserved productivity shock, and $\eta _{it}$ is, e.g.
measurement error in output that is often assumed to be uncorrelated across
time. \ For now we assume $\omega _{it}$ follows a first-order
autoregressive process
\begin{equation}
\omega _{it}=\rho _{\omega }^{0}\omega _{i,t-1}+\xi _{it},  \label{eq:x_w}
\end{equation}%
where $\xi _{it}$ is the innovation components of this process. A simple
identification restriction that has been used in the production function
literature involves assuming the following timing and information set
conditions.

\begin{assumption}
$E[\xi _{i,t+1}|I_{it}]=0$, where $I_{it}$ represents firm $i$'s information
set at $t$ and includes $x_{it}$ up to $t$ and $y_{it}$ up to $t-1$.
\end{assumption}

This is a simple form of a \textquotedblleft timing and information
set\textquotedblright\ assumption that has been used to address the
endogeneity of $x_{it}$ in (\ref{eq:pf0}). Loosely speaking, the assumption
can be interpreted as following from an economic model where (i) firms do
not observe $\omega _{it}$ until $t$, and (ii) firms endogenously choose $%
x_{it}$ at $t$ - thereby implying (with some additional, more technical
assumptions) that future innovations in $\omega _{it}$, i.e. $\xi _{i,t+1}$,
are uncorrelated with $x_{it}$.\footnote{%
Also note why this is not only a timing assumption, but also an assumption
about firms' information sets at various points in time. \ For example,
consider an alternative information set assumption where firms observe the
shocks $\omega _{it}$ \textit{one period ahead of time}, i.e. $\omega
_{i,t+1}$ is observed at $t$. \ Since, in general, firm choice of inputs $%
x_{it}$ may have dynamic implications (e.g. capital, labor with any sort of
adjustment costs), those choices could optimally depend on $\omega_{i,t+1}$,
implying that $\xi_{i,t+1}$ would no longer be uncorrelated with $x_{it}$.
On the other hand, suppose $\omega_{i,t+1}$ continues to be observed at $t$,
except with an alternative timing assumption where firms must commit to
their choice of $x_{it}$ at $t-1$. This case brings us back to $\xi_{i,t+1}$
being uncorrelated with $x_{it}$ once again.}

For estimation, one also needs to make an assumption on the measurement
error in output, typically $E[\eta _{it}|I_{it}]=0$. \ Given this, one can
consider estimating the model parameters $(\alpha ,\beta ,\rho _{\omega })$
by $\rho _{\omega }$-differencing the production function
\begin{eqnarray*}
y_{it}-\rho _{\omega }^{0}y_{i,t-1} &=&\alpha ^{0}(1-\rho _{\omega
}^{0})+\beta ^{0}(x_{it}-\rho _{\omega }^{0}x_{i,t-1})+\omega _{it}-\rho
_{\omega }^{0}\omega _{i,t-1}+\eta _{it}-\rho _{\omega }^{0}\eta _{i,t-1} \\
&=&\alpha ^{0}(1-\rho _{\omega }^{0})+\beta ^{0}(x_{it}-\rho _{\omega
}^{0}x_{i,t-1})+\xi _{it}+\eta _{it}-\rho _{\omega }^{0}\eta _{i,t-1}.
\end{eqnarray*}%
Since, by assumption, $\xi _{it}$, $\eta _{it}$, and $\eta _{i,t-1}$ are all
conditional mean independent of $I_{i,t-1}$, it follows that%
\begin{equation}
E[(y_{it}-\rho _{\omega }y_{i,t-1})-\alpha (1-\rho _{\omega })-\beta
(x_{it}-\rho _{\omega }x_{i,t-1})|I_{i,t-1}]=0\text{ \ \ \ \ at }(\alpha
,\beta ,\rho _{\omega })=(\alpha ^{0},\beta ^{0},\rho _{\omega }^{0}).
\label{eq:theequation}
\end{equation}

This is a very simple example of a type of ``dynamic panel'' moment
condition that has been used to estimate many production functions (see,
e.g., Blundell and Bond (2000) and Ackerberg, Caves, and Frazer (2015) for a
discussion of how the dynamic panel approach is related to proxy variable
approaches). Note that this model is simpler than those used in many actual
applications of dynamic panel methods. In particular, dynamic panel
approaches often include a fixed effect $\alpha _{i}$ in the production
function. This typically requires a second differencing to form usable
moment conditions.\footnote{
Moreover, some dynamic panel models include lagged $y$ (i.e. models with
state dependence) instead of the serially correlated $\omega$ in our model.
The nature of these alternative models is somewhat different. Including
lagged $y$ directly in the structural model means the persistence in $y$ is
causal (lagged $y$ causes current $y$), while in the models we study, the
persistence originates from the persistent unobserved state $\omega$. The
models we study (without lagged $y$) are used more often in the production
function literature\ (as well as the demand and hedonic literature cited in
the introduction), as there is typically more concern about serially
correlated unobserved productivity shocks than true state dependent
phenomena (although there are exceptions, e.g. a production function with
learning-by-doing has causal state dependence).} Our simplified model (\ref
{eq:theequation}) excludes fixed effects for expositional purposes. In a
moment we show that our under-identification result also applies to a model
with a fixed effect.

We know that the moment conditions (\ref{eq:theequation}) equal zero at the
true parameter values. \ But we now show that these moment conditions may
also equal zero at another set of parameter values. \ To study this, we need
to consider a complete model of the data generating process (DGP) for the
endogenously chosen $x_{it}$. \ Note that the typical paper in this
literature does not specify a full DGP for $x_{it}$ - this is because an
explicit DGP for $x_{it}$ is not needed to form the moments conditions (\ref%
{eq:theequation}) and attempt to estimate $(\alpha ,\beta ,\rho _{\omega })$%
. Specifying a form of the DGP for $x_{it}$ is important for understanding
the identification issue we study. As we will discuss in the next section,
this issue is ubiquitous under various forms of the DGP.

We start with what might be the simplest possible model of an endogenous $%
x_{it}$ that is correlated with productivity shock $\omega_{it}$:
\begin{equation}
x_{it}=\pi ^{0}+\theta ^{0}\omega _{it}+\kappa _{it}.  \label{eq:inputeq}
\end{equation}%
This allows a firm's choice of $x_{it}$ to depend on productivity shock $%
\omega _{it}$ as well as another persistent unobserved market factor $\kappa
_{it}.$ Analogous to $\omega _{it}$, we allow $\kappa _{it}$ to follow a
first-order autoregressive process
\begin{equation}
\kappa _{it}=\rho_{x}^{0}\kappa _{i,t-1}+u_{it},  \label{eq:x_ar}
\end{equation}%
where $u_{it}$ is the innovation component of the process. For example, if
$x_{it}$ measures labor input, then $\kappa _{it}$ can be interpreted as a
wage factor that is persistent over time.\footnote{Serial correlation in
productivity and/or input price shocks is widely
allowed in the literature. Empirical studies that find strong persistence in
firm input prices include Alonso-Borrego and Arellano (1999) and Grieco, Li,
and Zhang (2016). But it is worth noting that while we allow $\omega _{it}$
and $\kappa _{it}$ to be serially correlated, we do not require it.}
Analogous to how the innovation term in the $\omega _{it}$ process satisfies
$E[\xi _{i,t+1}|I_{it}]=0$, we assume that the innovation term in the $%
\kappa _{it}$ process satisfies $E[u_{i,t+1}|I_{it}]=0$. \ This is
consistent with a model in which firms observe $\omega _{it}$ and $\kappa
_{it}$ prior to choosing $x_{it}$, but who do not observe $\xi _{i,t+1}$ and
$u_{i,t+1}$ (and thus $\omega _{i,t+1}$ and $\kappa _{i,t+1}$) until the
next period.

We now show that with this DGP for $x_{it}$, the moment conditions (\ref
{eq:theequation}) generally have an \textit{additional zero} - one where
$\rho_{\omega}=\rho _{x}^{0}$, i.e. where the productivity persistence
parameter is not equal to its true value, but instead equal to the
persistence parameter of the \textit{input} process. This identification
issue has been noted in the context of the proxy variable literature on
estimating production functions (see footnote 16 of ACF and discussions in
Kim, Luo, and Su (2019)). KLS (2019) studies this pseudo-solution using
Monte Carlo simulations in the context of the proxy variable literature. \
In this paper, we show that this under-identification is a more general
issue that also applies to the dynamic panel literature. \ Studying the
issue in the dynamic panel approach is particularly useful because the
framework is simpler in some respects, and this allows us to better
understand the nature of the under-identification problem and to
propose more general approaches that we think might help mitigate the
problem.

Our derivation in the Appendix shows that, at this alternative parameter
vector $(\alpha,\beta,\rho )=(\alpha ^{0}-\frac{\pi^{0}}{\theta^{0}}%
,\beta^{0}+\frac{1}{\theta^{0}},\rho_{x}^{0})$, our simple dynamic panel
moment condition essentially recovers the innovation in the input demand
shock $\kappa _{it}$ (i.e. $u_{it}$) rather than the innovation in the
productivity shock $\omega_{it}$ (i.e. $\xi_{it}$). So given that we
assume innovation in $\kappa_{it}$ has the same properties as innovation in
$\omega_{it}$, our moment condition also equals zero at this
pseudo-solution. \ This is perhaps the simplest illustration of the
under-identification issue we study, though we develop more intuition for it
momentarily.\footnote{%
Note that this pseudo-solution still exists even if we include the
unobservables $\omega_{it}$ and $\kappa _{it}$ in the information set $%
I_{it}$. In general, the identified set may depend on the information
available in data and thus the number of periods observed. In practice, it
may also depend on the chosen instruments in the estimation.}

\begin{theorem}
Both $(\alpha ^{0},\beta ^{0},\rho _{\omega }^{0})$ and $(\alpha ^{0}-\frac{%
\pi ^{0}}{\theta ^{0}},\beta ^{0}+\frac{1}{\theta ^{0}},\rho _{x}^{0})$
solve the conditional moment condition \eqref{eq:theequation} given the
information set $I_{it}$.
\end{theorem}

Because the two zeros of the moment condition arise at different values of $%
\rho$, one may think that the identification issue does not arise when $%
\rho_{\omega }^{0}=\rho _{x}^{0}$. While this is not the focus of the paper,
we show that in this case the identification problem is even worse - in
particular, in this case there is an entire region of non-identification. To
see this argument from the input demand equation (\ref{eq:inputeq}), note
that if $\rho _{\omega }^{0}=\rho _{x}^{0}=\rho ^{0}$, then
\begin{align}
x_{it}-\rho ^{0}x_{i,t-1}& =\pi ^{0}(1-\rho ^{0})+\theta ^{0}(\omega
_{it}-\rho ^{0}\omega _{i,t-1})+(\kappa _{it}-\rho ^{0}\kappa _{i,t-1})
\notag \\
& =\pi ^{0}(1-\rho ^{0})+\theta ^{0}\xi _{it}+u_{it}.  \notag
\end{align}%
In other words, when $\rho _{\omega }^{0}=\rho _{x}^{0}$, the differencing $%
x_{it}-\rho x_{i,t-1}$ dispenses with all remaining information in $x_{it}$
other than the innovation terms, which are orthogonal to $I_{i,t-1}$. It
follows that, for an arbitrary $(\alpha ,\beta )$, moment equation (\ref%
{eq:theequation}) at $(\alpha ,\beta ,\rho _{\omega })=(\alpha ,\beta ,\rho
^{0})$ satisfies
\begin{align*}
0=& E[(y_{it}-\rho ^{0}y_{i,t-1})-\alpha (1-\rho ^{0})-\beta (x_{it}-\rho
^{0}x_{i,t-1})|I_{i,t-1}] \\
=& E[(\alpha ^{0}-\alpha )(1-\rho ^{0})+(\beta ^{0}-\beta )(x_{it}-\rho
^{0}x_{i,t-1})+\xi _{it}+\eta _{it}-\rho ^{0}\eta _{i,t-1}|I_{i,t-1}] \\
=& E[(\alpha ^{0}-\alpha )(1-\rho ^{0})+(\beta ^{0}-\beta )(\pi ^{0}(1-\rho
^{0})+\theta ^{0}\xi _{it}+u_{it})+\xi _{it}+\eta _{it}-\rho ^{0}\eta
_{i,t-1}|I_{i,t-1}] \\
=& (\alpha ^{0}-\alpha )(1-\rho ^{0})+(\beta ^{0}-\beta )[\pi ^{0}(1-\rho
^{0})],
\end{align*}%
where the second and third equalities come from substituting in the true DGP
from (\ref{eq:pf0}) and (\ref{eq:inputeq}), respectively, for which we
impose $\rho _{\omega }^{0}=\rho _{x}^{0}=\rho ^{0}$. The last result holds
since, given the timing and information set assumptions, the unobservables $%
\xi _{it}$, $u_{it}$, $\eta _{it}$, and $\eta _{i,t-1}$ are orthogonal to
the information available at time $t-1$. In other words, the moment equation
has arbitrarily many solutions of $(\alpha ,\beta )$ that satisfy
\begin{equation*}
(\alpha ^{0}-\alpha )(1-\rho ^{0})+(\beta ^{0}-\beta )[\pi ^{0}(1-\rho
^{0})]=0.
\end{equation*}%
Therefore, moment condition (\ref{eq:theequation}) does not contain any
information on $\beta $ when $\rho _{\omega }^{0}=\rho _{x}^{0}$.
Intuitively, this is similar to the non-identification result for
time-invariant variables in a fixed effects panel model. In any case, our
focus is on the existence of the pseudo-solution that arises from moment
condition (\ref{eq:theequation}) in a general setting. Therefore, for the
remainder of the paper, we avoid this more severe problem by assuming:

\begin{assumption}
\label{ineqrho} $\rho _{\omega }^{0}\neq \rho _{x}^{0}$,
\end{assumption}

\noindent which, again, is analogous to ruling out a time-invariant $x_{it}$
in a fixed effects model. In fact, this assumption has testable implications.\footnote{
One can show that $x_t$ has an AR(2) reduced form under Assumption \ref{ineqrho}
but an AR(1) reduced form otherwise. See the previous
version of this paper for more details.}

Returning to our main point, when $\rho_{\omega }^{0}\neq \rho _{x}^{0}$ we
have verified that the basic moment condition of the dynamic panel approach
may fail identification by yielding two solutions: one at the true AR(1)
parameter of the productivity shock and the other at the AR(1) parameter of
the input demand shock. It is important to note that this
identification problem is an issue with moment condition
(\ref{eq:theequation}). It is not necessarily a problem of the full
model, i.e. there is not necessarily an identification problem if one were
to impose all the restrictions of the model, e.g. those relating to higher
order moments.  In fact, Hu and Shum (2012) use deconvolution techniques to
show non-parametric identification of a first-order Markov model that is
similar to what we study.  However, their results rely on higher order
moments.  For example, to identify a first-order Markov model, Hu and Shum (2012)
require more than 2 time periods (in some cases up to 5 time periods) of
data, i.e. they utilize the strong restrictions the first-order Markov
assumption places on data in non-consecutive time periods. In our view,
even though the full model might be identified, it is important to study
identification problems based on (\ref{eq:theequation}) because the
vast majority of the empirical literature using timing and information set
assumptions (e.g. that following Olley and Pakes (1996) and Blundell and
Bond (2000)) utilizes conditional moments similar to
(\ref{eq:theequation}).

Momentarily we consider a possible solution to this problem that utilizes
additional moments based on the DGP of $x_{it}$, benefiting from the fact
that, in the production function context (and likely in many other
contexts), one may be willing to make assumptions about the sign of $\theta $%
, i.e. the direction of endogeneity. But first we examine the extent to
which this pseudo-solution problem manifests itself in more general models.

\section{Pseudo-Solutions in More General Models}

The above analysis identified a pseudo-solution in the context of a
simplified dynamic panel model, combined with an assumed DGP for $x_{it}$
that was also very simple. A natural question to ask is to what extent this
pseudo-solution exists in more general dynamic panel models. Is this
pseudo-solution a ``knife-edged'' phenomenon that only occurs with a very
specific production function or productivity process, or that only occurs in
models where the productivity and the factor input shocks have the same
structure (e.g. both AR(1))? Or is it a more general phenomenon?

To address these questions, we now consider different or more general models
- either in terms of the structural function of interest, the unobserved
productivity or input shock processes, or the timing assumptions of the
model. While we are only able to derive theoretical results for a subset of
these alternative models, in other models we use simulations to study their
objective functions (with very large datasets to approximate their
asymptotic values).

In some of our examples we can show that the exact same pseudo-solution
identified above persists in the more general model we study. In others,
there is still a pseudo-solution but at a different (though ``nearby'')
point in parameter space. In others, there are multiple pseudo-solutions
that we can locate. In still others, the pseudo-solution that sets the
moments exactly equal to zero disappears, but there is a nearby parameter
that sets the moments ``close-to-zero'' and generates a local minimum of the
GMM objective function. While in this last case the model is formally
identified, we still think it important to note because the existence of
local minima can generate challenges in numeric optimization (and the
guidance we provide later in the paper could also help overcome these
numeric challenges). Notably, in one of our extensions - one where we
strengthen the timing assumptions of the model - the pseudo-solution appears
to completely disappear.

\subsection{Extended Model with Fixed Effects}
Traditional dynamic panel models also include fixed effects in the main
equation, e.g.
\begin{equation}
y_{it}=\alpha _{i}+\beta ^{0}x_{it}+\omega _{it}+\eta _{it}.  \label{eq:fe}
\end{equation}%
Here, we show that this model also exhibits the pseudo-solution identified
above. \

In model (\ref{eq:fe}), forming usable moments requires an additional
difference to remove the fixed effect $\alpha _{i}$. \ More specifically,
the $\rho _{\omega }^{0}$-difference
\begin{equation*}
y_{it}-\rho _{\omega }^{0}y_{i,t-1}=\alpha _{i}(1-\rho _{\omega }^{0})+\beta
^{0}(x_{it}-\rho _{\omega }^{0}x_{i,t-1})+\xi _{it}+\eta _{it}-\rho _{\omega
}^{0}\eta _{i,t-1}
\end{equation*}%
contains $\alpha _{i}$, so an additional first difference is typically done
to obtain%
\begin{eqnarray*}
&&\left( y_{it}-\rho _{\omega }^{0}y_{i,t-1}\right) -\left( y_{i,t-1}-\rho
_{\omega }^{0}y_{i,t-2}\right) \\
&=&\beta ^{0}\left( (x_{it}-\rho _{\omega }^{0}x_{i,t-1})-(x_{i,t-1}-\rho
_{\omega }^{0}x_{i,t-2})\right) +\xi _{it}-\xi _{i,t-1}+\left( \eta
_{it}-\rho _{\omega }^{0}\eta _{i,t-1}\right) -\left( \eta _{i,t-1}-\rho
_{\omega }^{0}\eta _{i,t-2}\right).
\end{eqnarray*}%
Since this eliminates $\alpha _{i}$, one can proceed to estimate $\beta ^{0}$
with a moment condition based on this, e.g.
\begin{equation}
E[\left( y_{it}-\rho _{\omega }y_{i,t-1}\right) -\left( y_{i,t-1}-\rho
_{\omega }y_{i,t-2}\right) -\beta \left( (x_{it}-\rho _{\omega
}x_{i,t-1})-(x_{i,t-1}-\rho _{\omega }x_{i,t-2})\right) |I_{i,t-2}]=0.
\label{femom1}
\end{equation}

To show that this moment condition may also have the under-identification
problem, again consider a DGP for input $x_{it}$. \ Since the DGP for $%
y_{it} $ includes a fixed effect, it is natural to include one here as well,
e.g.
\begin{equation}
x_{it}=\pi _{i}+\theta ^{0}\omega _{it}+\kappa _{it}.  \notag
\end{equation}%
Following the previous argument, substitute the inverted production function
into
\begin{eqnarray*}
x_{it} &=&\pi _{i}+\theta ^{0}\left( y_{it}-\alpha _{i}-\beta
^{0}x_{it}-\eta _{it}\right) +\kappa _{it} \\
&=&\left( \frac{1}{\beta ^{0}\theta ^{0}+1}\right) \left( \pi _{i}-\theta
^{0}\alpha _{i}+\theta ^{0}y_{it}-\theta ^{0}\eta _{it}+\kappa _{it}\right)
\end{eqnarray*} and $\rho _{x}^{0}$-difference to obtain
\begin{equation*}
x_{it}-\rho _{x}^{0}x_{i,t-1}=\left( \frac{1}{\beta ^{0}\theta ^{0}+1}%
\right) \left( \left( 1-\rho _{x}^{0}\right) \left( \pi _{i}-\theta
^{0}\alpha _{i}\right) +\theta ^{0}\left( y_{it}-\rho
_{x}^{0}y_{i,t-1}\right) -\theta ^{0}\left( \eta _{it}-\rho _{x}^{0}\eta
_{i,t-1}\right) +u_{it}\right).
\end{equation*}
First differencing this equation results in
\begin{multline*}
(x_{it}-\rho_{x}^{0}x_{i,t-1}) -(x_{i,t-1}-\rho
_{x}^{0}x_{i,t-2}) = (\frac{\theta^{0}}{\beta ^{0}\theta ^{0}+1})((
y_{it}-\rho _{x}^{0}y_{i,t-1})-(y_{i,t-1}-\rho
_{x}^{0}y_{i,t-2}))\\
-(\frac{\theta^{0}}{\beta^{0}\theta ^{0}+1})((
\eta _{it}-\rho_{x}^{0}\eta _{i,t-1}) -(\eta_{i,t-1}-\rho
_{x}^{0}\eta_{i,t-2})) + (\frac{1}{\beta^{0}\theta^{0}+1})(u_{it}-u_{i,t-1}),
\end{multline*}
and given that the unobservable terms in this equation are assumed to be
mean independent of $I_{i,t-2}$, this implies%
\begin{equation}
E[( y_{it}-\rho _{x}^{0}y_{i,t-1}) -( y_{i,t-1}-\rho
_{x}^{0}y_{i,t-2}) -(\beta^{0}+\frac{1}{\theta ^{0}})
((x_{it}-\rho_{x}^{0}x_{i,t-1})-(x_{i,t-1}-\rho
_{x}^{0}x_{i,t-2}))|I_{i,t-2}]=0. \label{femom2}
\end{equation}

Comparing (\ref{femom2}) to (\ref{femom1}), it is clear that there will be a
pseudo-solution to moment condition (\ref{femom1}) in the same location as
the model without a fixed effect, i.e. where $\rho _{\omega }=\rho _{x}^{0}$
and $\beta =\beta ^{0}+\frac{1}{\theta ^{0}}$. Note that this argument does
not depend on the existence of a fixed effect in the input demand equation,
i.e. there is a pseudo-solution even if $\pi _{i}=\pi$. \

\subsection{Multiple Endogenous Variables}
We next examine the under-identification problem where we have multiple
variable inputs in the production function context or, more generally,
multiple endogenous regressors. Consider a panel regression model with two
endogenous regressors $x$ and $z$,
\begin{equation}
y_{it}=\alpha ^{0}+\beta ^{0}x_{it}+\gamma ^{0}z_{it}+\omega _{it}+\eta
_{it},  \label{eq:pf}
\end{equation}%
where
\begin{align*}
x_{it}& =\pi _{x}^{0}+\theta _{\omega }^{0}\omega _{it}+\theta _{\kappa
}^{0}\kappa _{it}+\theta _{\wp }^{0}\wp _{it}, \\
z_{it}& =\pi _{z}^{0}+\delta _{\omega }^{0}\omega _{it}+\delta _{\kappa
}^{0}\kappa _{it}+\delta _{\wp }^{0}\wp _{it}.
\end{align*}%
Here, $\wp _{it}$ and $\kappa _{it}$ denote two persistent market factors --
such as input prices -- and they follow autoregressive processes $\wp
_{it}=\rho _{z}^{0}\wp _{i,t-1}+v_{it}$ and $\kappa _{it}=\rho
_{x}^{0}\kappa _{i,t-1}+u_{it}$, respectively. \ Note that both inputs
depend on both factors -- this is typically the case in practice as the
optimal choice of one variable input typically depends on the prices of all
variable inputs.

On the one hand, exploiting the AR(1) process of productivity $\omega
_{it}=\rho_{\omega}^0\omega _{i,t-1}+\xi_{it}$, and $\rho_{\omega}^0$%
-differencing leads to
\begin{equation*}
y_{it}-\rho_{\omega}^0y_{i,t-1} = \alpha^0 (1-\rho_\omega^0) + \beta^0
(x_{it} - \rho_{\omega}^0 x_{i,t-1}) + \gamma^0 (z_{it} - \rho_{\omega}^0
z_{i,t-1}) + \xi_{it}+(\eta_{it} - \rho_\omega^0 \eta_{i,t-1}),
\end{equation*}
which can be used for constructing the following moment condition, as a
generalization of moment condition \eqref{eq:theequation},
\begin{equation}
E[y_{it}-\rho_{\omega}y_{i,t-1} - \alpha (1-\rho_\omega) - \beta(x_{it} -
\rho_{\omega} x_{i,t-1}) - \gamma (z_{it} - \rho_{\omega} z_{i,t-1})|
I_{i,t-1}]=0.  \label{eq:genmoment}
\end{equation}

On the other hand, consider a linear combination of the two endogenous
inputs
\begin{equation*}
\delta _{\kappa }^{0}x_{it}-\theta _{\kappa }^{0}z_{it}=(\delta _{\kappa
}^{0}\pi _{x}^{0}-\theta _{\kappa }^{0}\pi _{z}^{0})+(\delta _{\kappa
}^{0}\theta _{\omega }^{0}-\theta _{\kappa }^{0}\delta _{\omega }^{0})\omega
_{it}+(\delta _{\kappa }^{0}\theta _{\wp }^{0}-\theta _{\kappa }^{0}\delta
_{\wp }^{0})\wp _{it},
\end{equation*}%
where the expression on the right-hand side follows from the two input
functions. This shows that we can write a linear combination of two inputs
as a linear function of two sources of persistence (one for productivity and
one for market factor) -- similar in structure to the one input in the
benchmark model. Solving for $\omega _{it}$ and plugging it into the output
equation \eqref{eq:pf} leads to a similar structure of the output function,
except that $\omega _{it}$ is replaced by $\wp _{it}$ and the coefficients
on inputs are different from the true production function parameters. Notice
that this is similar in structure to the alternative expression of moment condition 
(\ref{eq:theproof}) we derived in the Appendix for the benchmark model.

Therefore, following similar arguments as before, we can derive a spurious
solution of moment function (\ref{eq:genmoment}) by taking $\rho_z^0$%
-differencing due to the autoregressive process of $\wp_{it}$, instead of
the productivity. As the result, this spurious solution is at the dynamic
parameter $\rho_\omega=\rho_z^0$ of the persistence market factor $\wp_{it}$%
, not the productivity. Similarly, a different linear combination $%
\delta_\wp^0 x_{it} - \theta_\wp^0 z_{it}$ leads to another spurious
solution of moment (\ref{eq:genmoment}). This solution is at the dynamic
parameter $\rho_\omega=\rho_x^0$ of the persistence market factor $%
\kappa_{it}$.

In summary, we conclude that when the production function has multiple
inputs, the dynamic panel moments may yield the same number of spurious
solutions as the number of flexible inputs that have persistent shocks.

\subsection{Dynamic Input}

The under-identification problem also still exists when we have both
flexible and dynamic inputs. Consider a panel regression model with a
flexible input $x$ and a dynamic input $z$ like capital as follows
\begin{equation}
y_{it}=\alpha ^{0}+\beta ^{0}x_{it}+\gamma ^{0}z_{it}+\omega _{it}+\eta
_{it},  \label{eq:fixed}
\end{equation}%
where the flexible input function is given by
\begin{equation*}
x_{it}=\pi ^{0}+\theta _{\omega }^{0}\omega _{it}+\theta
_{z}^{0}z_{it}+\kappa _{it},
\end{equation*}%
and persistent market factor $\kappa _{it}$ follows the AR(1) process $%
\kappa _{it}=\rho _{x}^{0}\kappa _{i,t-1}+u_{it}$ as before.

Solving for $\omega _{it}$ in the input function and plugging it into %
\eqref{eq:fixed} leads to a similar structure of the alternative expression
of the output function, except that $\omega _{it}$ is replaced by $\kappa
_{it}$ and the coefficients on inputs are different from the true production
function. Following similar arguments as before, we can derive a spurious
solution of the dynamic panel moment due to this similar structure where the
spurious solution is at the dynamic parameter $\rho _{\omega }=\rho _{x}^{0}$
of the persistent input factor $\kappa _{it}$. Note that
we do not expect a spurious identification problem with respect to the
dynamic input $z_{it}$. In other words, if $\beta^0 = 0$ in the above model so there
is only a dynamic input $z_{it}$, we do not expect a spurious solution.  This
is because the input demand function for a dynamic input like $z_{it}$ will
depend on factors other than the shocks, e.g. $z_{it-1}$ and/or further past
$z_{it}$'s.

\subsection{Beyond Linearity and AR(1)}
In the above three extensions to our basic model, we were able to
analytically show that the pseudo-solution (or pseudo-solutions) continues
to be a potential problem. In this section, we consider some additional
extensions. While we are unable to analytically show the existence of
pseudo-solutions in these extensions, we are able to use simulations to show
that they often continue to exist, at least in neighborhoods of our
benchmark model. In particular, we consider relaxing the assumed linearity
of the input demand function and the AR(1) process for unobserved shock $%
\kappa _{it}$ in the input demand function.

It is important to note that we continue to assume that $\omega _{it}$
follows an AR(1) process. This means that our basic dynamic panel moment
condition (\ref{eq:theequation}) continues to be zero at the true parameter
value. \ What we are focused on is not the zero at the true parameter value,
but whether there continues to be a pseudo-solution (or pseudo-solutions) at
other parameters, and the location of such pseudo-solutions. \ There is no
need to do multiple Monte Carlo replications to study this -- it is easier
to simply boost the number of observations (to 200,000) to simulate the
``asymptotic'' objective function and examine the location of zeros. In the
benchmark case we know there are ``asymptotic'' zeros at the true parameter,
set to $\beta=\beta^0 =0.6$, and at the pseudo-parameter $\beta
=\beta^0+1/\theta^0=0.6+1/1=1.6$ (since in our benchmark model, we set $%
\theta^0=1$).\footnote{Our benchmark case for these simulations uses 
the same parameter values as
used in the Monte-Carlo experiments, see Section 5 for more details.} In the
alternative models, we know there is an ``asymptotic'' zero at the true
parameter $\beta =0.6$. The question is whether there are pseudo-solutions -
either at $\beta =1.6$ or elsewhere.

\begin{figure}[!h]
\caption{Nonlinearity in the Productivity Process}
\label{prodnonlinear}\centering
\includegraphics[width=1\textwidth]{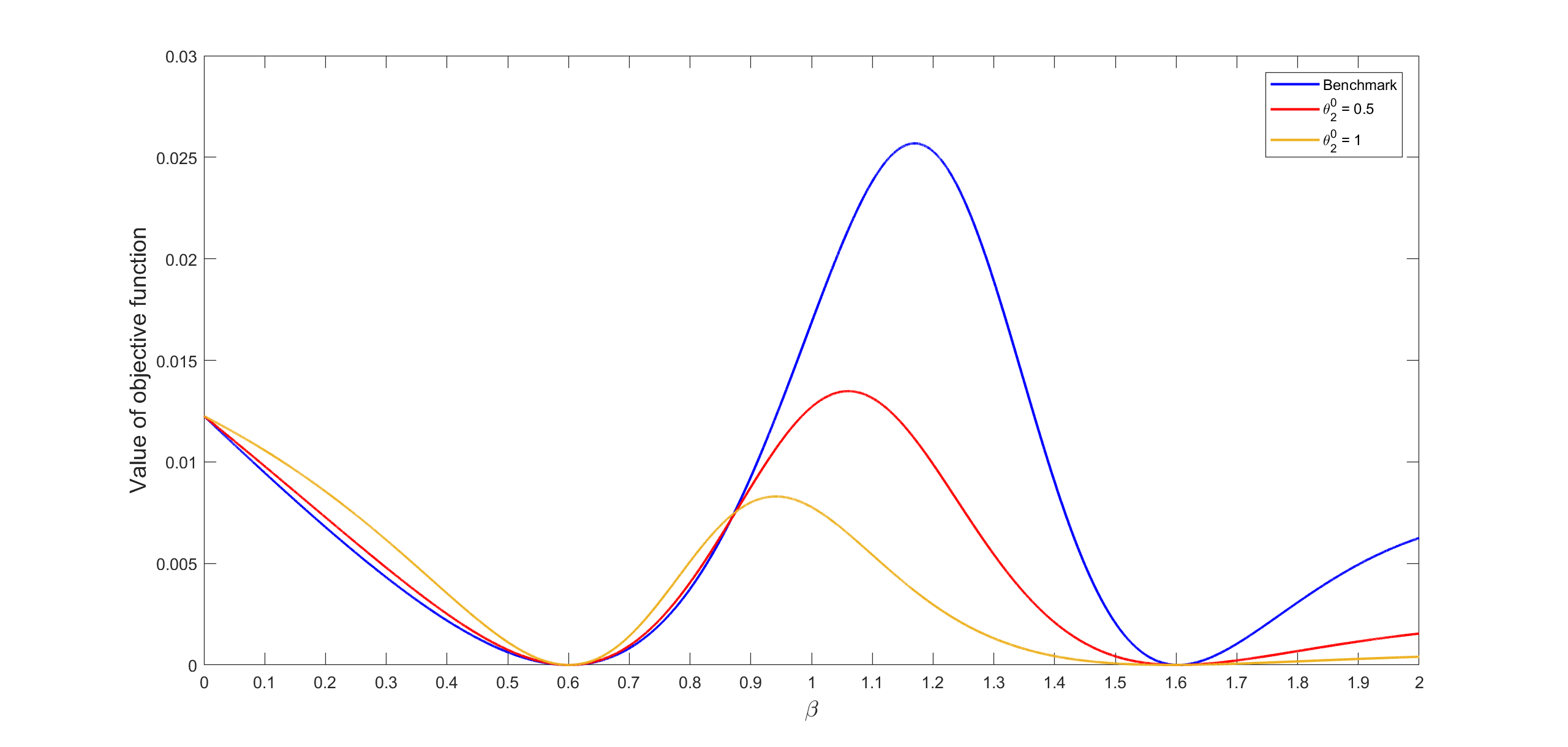}
\end{figure}

\begin{figure}[!h]
\caption{Nonlinearity in the Input Market Persistence Process}
\label{kappanonlinear}\centering
\includegraphics[width=1\textwidth]{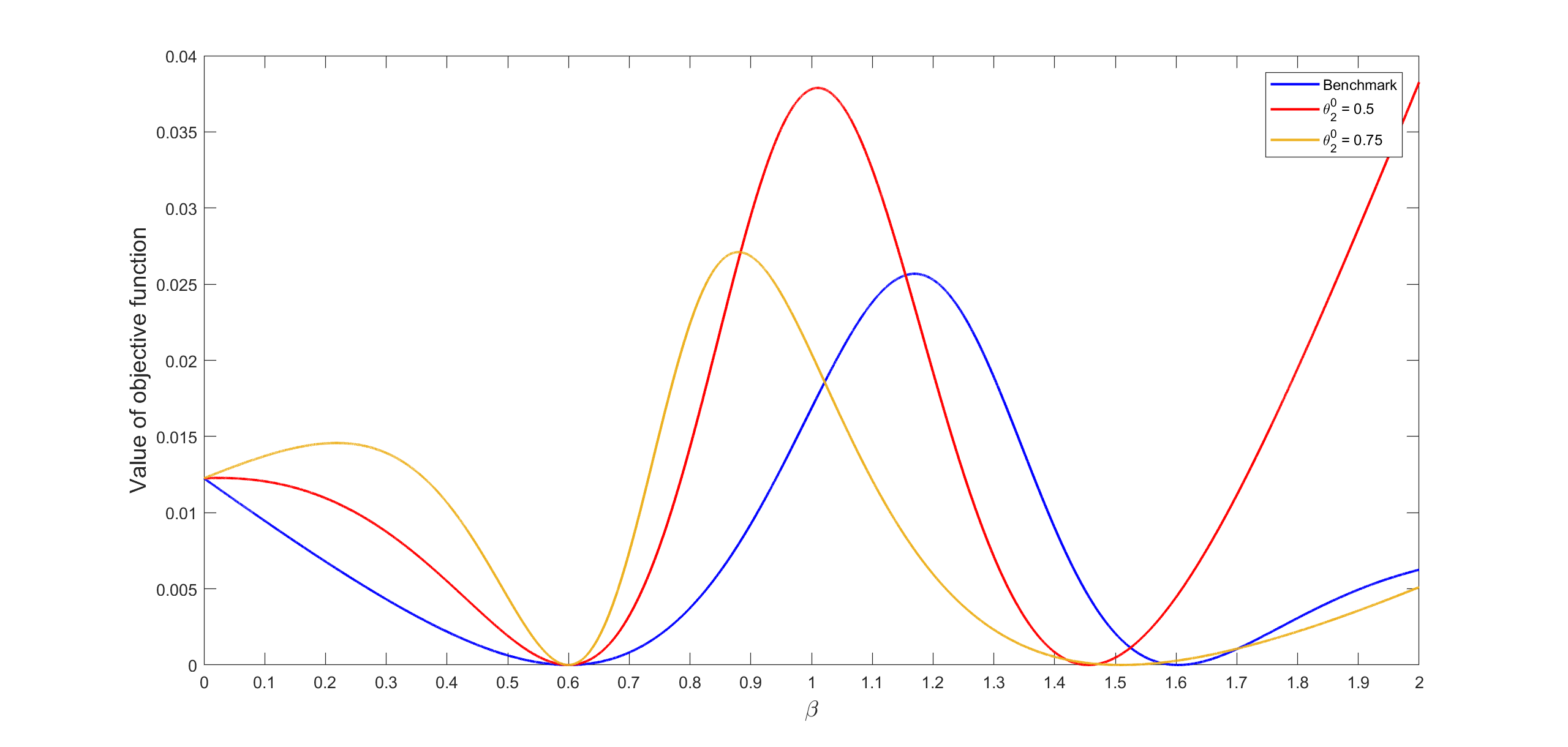}
\end{figure}

\begin{figure}[!h]
\caption{AR(2)}
\label{kappaar2}\centering
\includegraphics[width=1\textwidth]{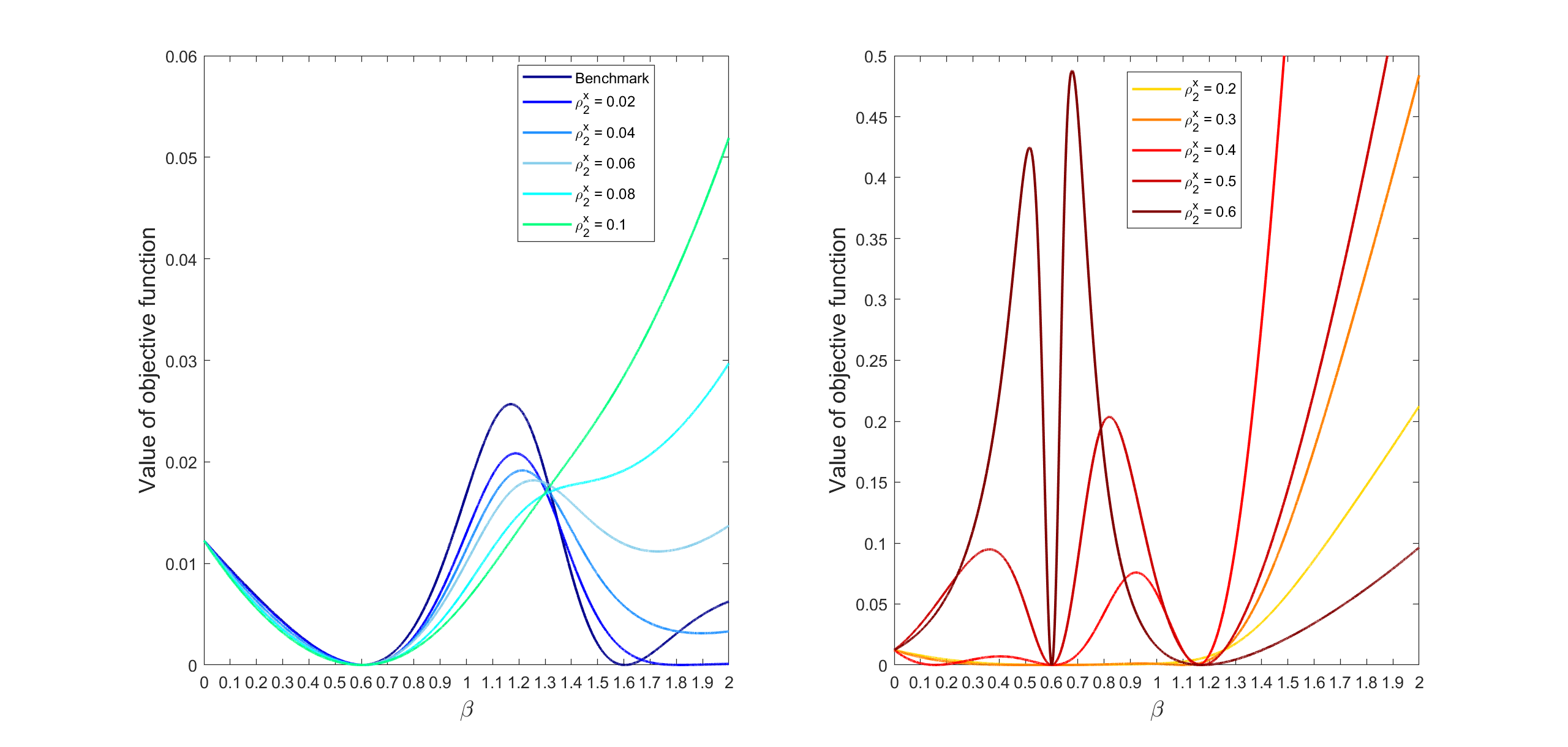}
\end{figure}

\begin{figure}[!h]
\caption{Noisy Input Process}
\label{xshock}\centering
\includegraphics[width=1\textwidth]{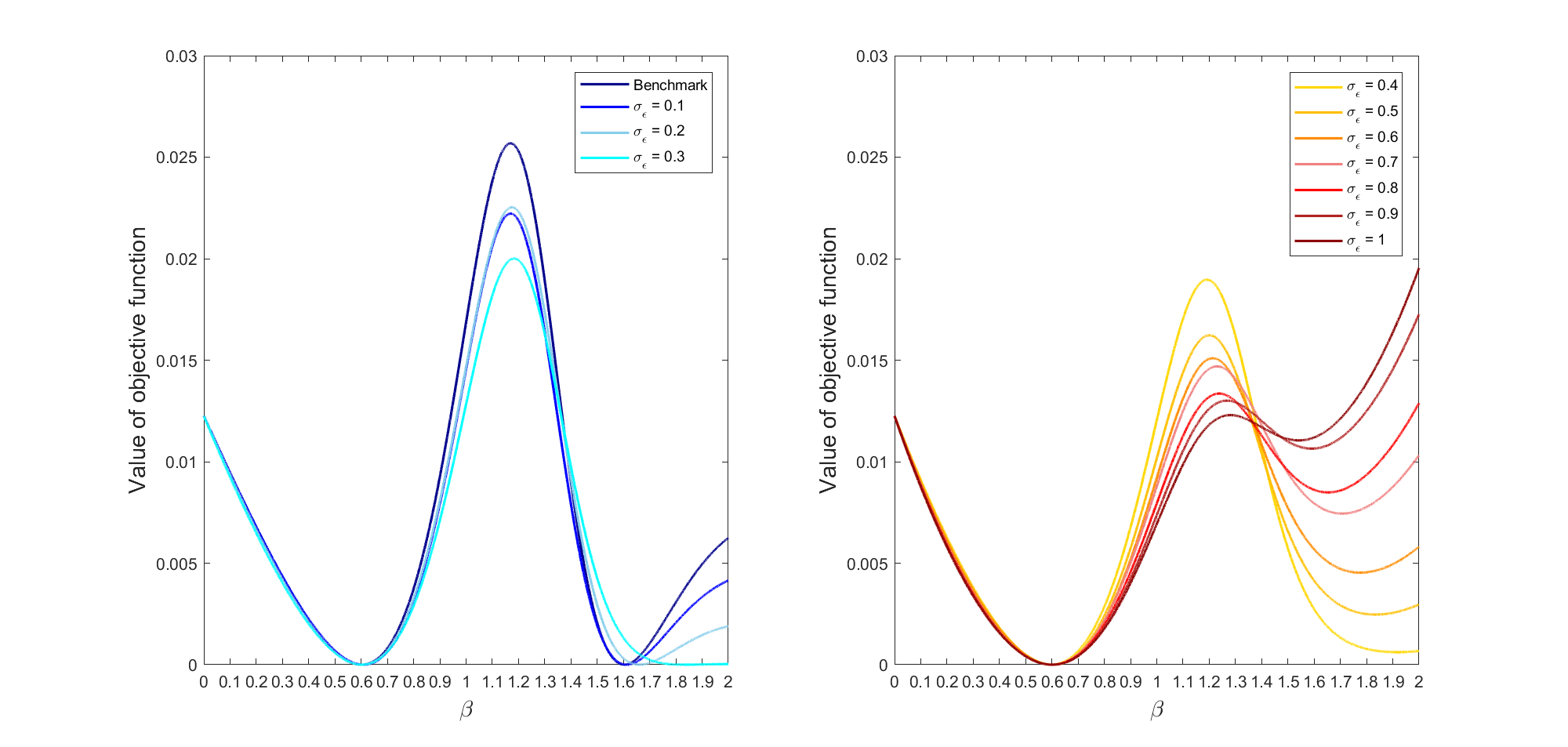}
\end{figure}

\begin{figure}[!h]
\caption{Alternative Model Where the Curvature of the Non-linearity in $%
\protect\kappa _{it}$ is Reversed}
\label{alternative}\centering
\includegraphics[width=1\textwidth]{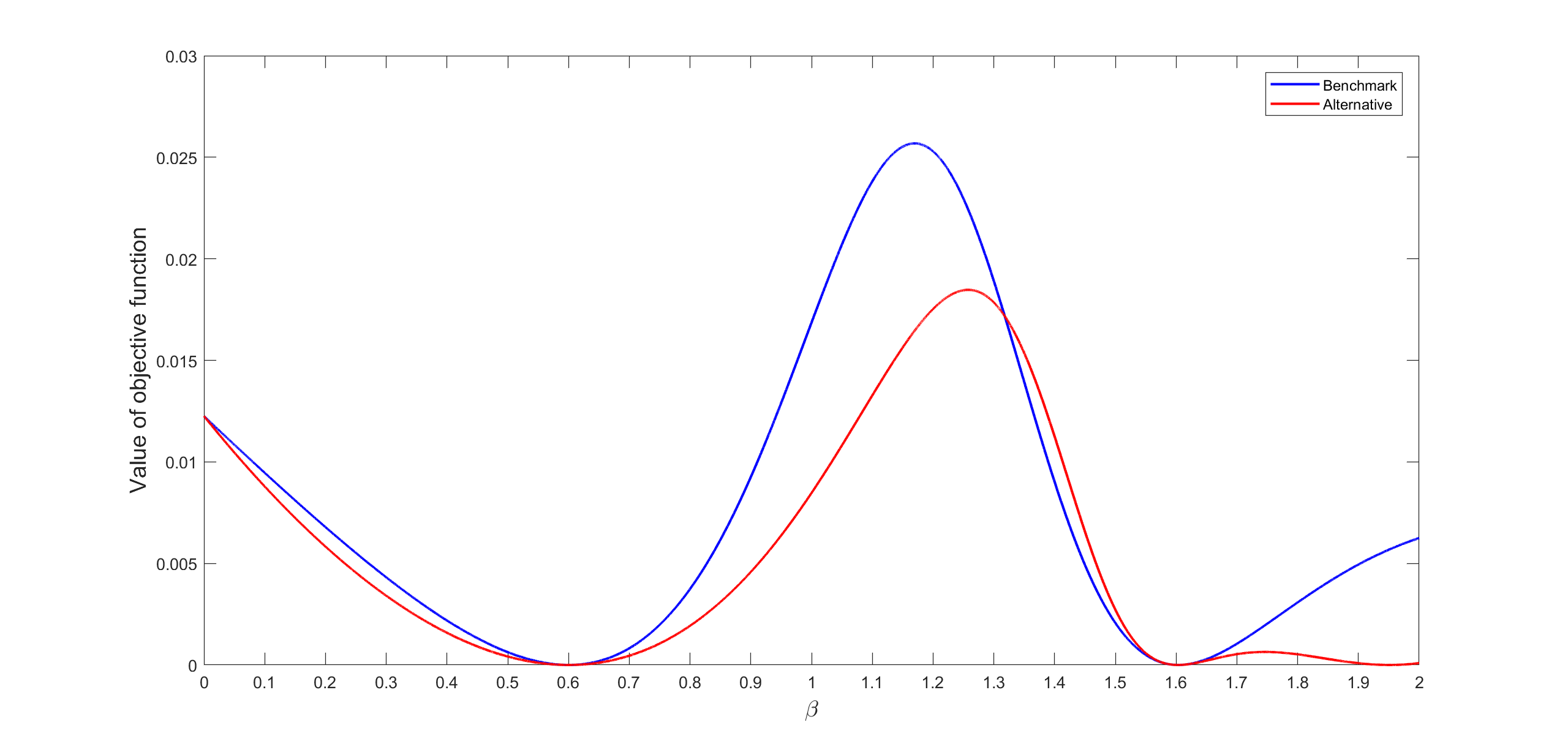}
\end{figure}

First, we consider an alternative model where the input function is a
non-linear function of the productivity process, i.e.:
\begin{equation}
x_{it}=\pi ^{0}+\theta _{1}^{0}\omega _{it}+\theta _{2}^{0}\omega
_{it}^{2}+\kappa _{it},  \notag
\end{equation}%
where we set $\theta _{1}^{0}$ = $1$, as in the benchmark model, and vary $%
\theta _{2}^{0}$. Figure \ref{prodnonlinear} shows the basic dynamic panel
objective function (based on the moment condition (\ref{eq:theequation})),
for both the benchmark model (without nonlinearity, i.e. $\theta_{2}^{0}=0$)
and this alternative model where we set $\theta _{2}^{0}=0.5$ and $1$.\footnote{
For ease of comparison, we rescale objective functions such that they are
equal at $\beta =0$ in Figures \ref{prodnonlinear}-\ref{alternative}.} For
easy visualization, we display the objective function as a function of just $
\beta$ by concentrating out the constant term and AR(1) parameter.\footnote{
To concentrate out $\rho_{\omega}$ and the constant, for a candidate $\beta
=\widetilde{\beta}$, we estimate $\rho_{\omega}$ and the constant with the
linear equation
\begin{equation*}
\left( y_{it}-\widetilde{\beta }x_{it}\right) =\alpha (1-\rho _{\omega
})+\rho _{\omega }\left( y_{i,t-1}-\widetilde{\beta }x_{i,t-1}\right) +\xi
_{it}+\eta _{it}-\rho _{\omega }\eta _{i,t-1}
\end{equation*}
using instrumental variables with $(y_{i,t-2}-\widetilde{\beta}
x_{i,t-2})$ as an instrument for $(y_{i,t-1}-\widetilde{\beta}
x_{i,t-1}).$ So that this objective function illustrates exact zeros
(rather than ``close to zeros''), we use only $x_{i,t-1}$ as the instrument
for the concentrated moment condition for $\beta$.} The objective functions
for the alternative model also have pseudo-solutions, at $\beta$ quite close
to the pseudo-solution in the benchmark model. Interestingly, to the right
of the pseudo-solution, the objective function becomes flatter than the
benchmark case, especially when $\theta_{2}^{0}=1$. However, the main
finding is that the pseudo-solution does not appear to disappear when the $%
x_{it}$ process is not linear in $\omega _{it}$.

Second, we consider a model where there is a nonlinearity in the first-order
Markov process for $\kappa _{it}$. Specifically, we consider:
\begin{equation}
\kappa _{it}=\frac{e^{\theta _{2}^{0}\left\vert \kappa _{i,t-1}\right\vert }%
}{1+e^{\theta _{2}^{0}\left\vert \kappa _{i,t-1}\right\vert }}\kappa
_{i,t-1}+u_{it}.  \notag
\end{equation}%
When we set $\theta _{2}^{0}=0$, $\frac{e^{\theta _{2}^{0}\left\vert \kappa
_{i,t-1}\right\vert }}{1+e^{\theta _{2}^{0}\left\vert \kappa
_{i,t-1}\right\vert }}=0.5$ $\forall \kappa _{i,t-1}$, so this model is
equivalent to the benchmark case (where we set the AR(1)\ parameter = $0.5$%
). When $\theta_{2}^{0}$ $>0$, the Markov process is non-linear. One
interpretation of this is that the extent to which $\kappa _{i,t-1}$
depreciates depends on the level of $\left\vert \kappa _{i,t-1}\right\vert $%
. Note that since $\left\vert \kappa _{i,t-1}\right\vert $ is positive, the
heterogeneous depreciation rate in this model is between 0.5 and 1.\ Figure %
\ref{kappanonlinear} shows the (concentrated) objective function for this
model when $\theta _{2}^{0}=0.5$ and $0.75$. Again, the pseudo-solution
persists, but moves down relative to the benchmark model. In these examples,
the pseudo-solutions are at approximately 1.46 and 1.51 for $\theta
_{2}^{0}=0.5$ and $0.75$, respectively. Interestingly, the effect of $\theta
_{2}^{0}$ on the location of the pseudo-parameter is non-monotonic. This
appears to be because changing $\theta _{2}^{0}$ affects multiple things in
the model - e.g., it simultaneously changes both the average level of
depreciation and its heterogeneity.

Third, we change the $\kappa _{it}$ process in a different way - keeping
linearity but allowing it to follow an AR(2)\ process instead of an AR(1)\
process, i.e.
\begin{equation}
\kappa _{it}=\rho _{1}^{x}\kappa _{i,t-1}+\rho _{2}^{x}\kappa
_{i,t-2}+u_{it}.  \notag
\end{equation}%
Starting with{\ the benchmark case ($\rho _{1}^{x}=0.5$ and $\rho _{2}^{x}=0$%
) we consider alternative models where $\rho _{2}^{x}$ is set to various
values between 0 and 1. Figure \ref{kappaar2} shows the (concentrated)
objective functions and indicates that the existence of the pseudo-solution
may depend on the value of $\rho _{2}^{x}$ in this alternative model. When $%
\rho _{2}^{x}$ is close to 0, i.e. the $\kappa _{it}$ process is close to an
AR(1) model, the pseudo-solution persists, as shown in the left panel of
Figure \ref{kappaar2} for $\rho _{2}^{x}=0.02$. When we gradually increase $%
\rho _{2}^{x}$, the pseudo-solution disappears from the region. However, the
right panel suggests distinct patterns when we increase $\rho _{2}^{x}$
further. For $\rho _{2}^{x}=0.2$ and $0.3$, we spot a wide flat region where
the truth and the potential pseudo-solutions seem to be indistinguishable.
When $\rho _{2}^{x}$ becomes large enough, i.e. $\rho _{2}^{x}\geq 0.4$, a
pseudo-solution reappears but at a lower $\beta $, at approximately 1.17.
And there is even an additional pseudo-solution when $\rho _{2}^{x}=0.4$ to
the left of the truth, at approximately 0.16. Therefore, the AR(2) process
may complicate the identification of the model but for a wide range of $\rho
_{2}^{x}$, the pseudo-solution is still a potential concern.}

Lastly, we consider adding an additional i.i.d shock to the input process
(in addition to the AR(1) $\kappa _{it}$). In other words, we assume the
following input process
\begin{equation}
x_{it}=\pi ^{0}+\theta ^{0}\omega _{it}+\kappa _{it}+\epsilon _{it},
\label{eq:inputeq1}
\end{equation}%
where $\epsilon _{it}$ is i.i.d over time (and distributed normally in our
simulations). Essentially this is a specification where the input process
(conditional on $\omega _{it}$) follows a particular type of ARMA process.
\ With this process it is straightforward to show theoretically that at the
pseudo-parameters $(\alpha^{0}-\frac{\pi ^{0}}{\theta ^{0}},\beta^{0}+%
\frac{1}{\theta^{0}},\rho_{x}^{0})$, the dynamic panel moment condition is
non-zero - one can simply substitute the inverted input process into the
production function to get
\begin{equation*}
y_{it}=\alpha ^{0}-\frac{\pi ^{0}}{\theta ^{0}}+\left( \beta ^{0}+\frac{1}{%
\theta ^{0}}\right) x_{it}-\frac{\kappa _{it}}{\theta ^{0}}+\eta _{it}-\frac{%
\epsilon _{it}}{\theta ^{0}},
\end{equation*}%
and $\rho _{x}^{0}$ difference, resulting in
\begin{equation*}
y_{it}-\rho _{x}^{0}y_{i,t-1}=( \alpha ^{0}-\frac{\pi ^{0}}{\theta ^{0}})
(1-\rho _{x}^{0})+( \beta ^{0}+\frac{1}{\theta ^{0}})
(x_{it}-\rho _{x}^{0}x_{i,t-1})+\tau _{it},
\end{equation*}%
where $\tau _{it}=-\frac{1}{\theta ^{0}}u_{it}+(\eta _{it}-\rho
_{x}^{0}\eta _{i,t-1}) -\frac{1}{\theta ^{0}}(\epsilon
_{it}-\rho_{x}^0\epsilon _{i,t-1})$. Because of the additional $
\epsilon _{it}$ terms (and since $E[\epsilon _{i,t-1}|x_{i,t-1}]\neq 0$), $
E[\tau _{it}|I_{i,t-1}]\neq 0$ (unless $\rho _{x}^{0}\neq 0$). \ Thus, the
moment condition is non-zero at the pseudo-parameters identified above ($
\beta =1.6$ with the parameters used in our simulations).

Of course, this is just one point in parameter space. One might be concerned
that, as in the prior examples, there might still be a pseudo-solution, just
at a different parameter value. \ Simulations show that this does not always
appear to be the case. Figure \ref{xshock} shows the (concentrated)
objective function for the benchmark case ($\epsilon $ = 0) and in models
that add $\epsilon _{it}$ to the input process with different standard
deviations, i.e. $\sigma _{\epsilon }$ is set to between 0.1 and 1 with a
stepsize equal to 0.1. Interestingly, the left panel shows that when the
variance of the shock is small, a pseudo-solution appears to persist -
increasing somewhat from the $\beta =1.6$ location in the benchmark case. On
the other hand, when we further increase $\sigma _{\epsilon }$, the
pseudo-solution appears to disappear from the region, as suggested by the
right panel of Figure \ref{xshock}. Nonetheless, around the location of the
baseline pseudo-solution there is still a local minimum of the objective
functions in these models with larger $\sigma _{\epsilon }$. In practice,
these local minima like this can also be problematic for numeric
optimization, so it does not completely eliminate the potential issue.

In summary, these models suggest that the existence of pseudo-solutions is
not a knife-edge result that relies on the very specific input demand
equation in our benchmark model. While the exact locations of these
pseudo-solutions (or local minima in some cases) differ from the benchmark
model, the general patterns are similar - in particular, the
pseudo-solutions are typically at a larger $\beta $ than the true parameter $%
\beta ^{0}=0.6$. Obviously, this is only a small set of alternative models
and there could very well be other alternative models without
pseudo-solutions, especially as one moves parameters further away from the
benchmark model. In other experiments, we have found that many things can
happen - for example, in the following model (where the curvature of the
non-linearity in $\kappa _{it}$ is reversed
\begin{equation*}
\kappa _{it}=\left( 1-\frac{e^{\left\vert \kappa _{i,t-1}\right\vert }}{%
1+e^{\left\vert \kappa _{i,t-1}\right\vert }}\right) \kappa _{i,t-1}+u_{it},
\end{equation*}%
we see multiple pseudo-solutions, one slightly above 1.6, and one at
approximately $\beta =1.95$ (Figure \ref{alternative}). \ But overall, the
results hint that this is a general issue - the $\rho $-differencing moment %
\eqref{eq:theequation} of the dynamic panel approach (and likely of related
quasi-differencing approaches, e.g. the proxy variable approaches of
OP/LP/ACF) creates a quadratic form indeterminacy that seems to exist
relatively generally.

\subsection{Stronger Timing/Information Set Assumptions}

\label{sec:stronger}

Interestingly, we have found one class of modeling assumptions that does
appear to remove the pseudo-solution very generally. \ Recall Assumption 1,
the timing and information set assumption that the moment conditions of our
benchmark model are predicated on. \ Again, this can be interpreted as
following from an economic model where 1) firms do not observe $\omega _{it}$
and $\kappa _{it}$ until $t$, and 2) firms choose $x_{it}$ at $t$ and hence
implying (with some additional assumptions) that future innovations in $%
\omega _{it}$ and $\kappa _{it}$, i.e. $\xi _{i\tau }$ and $u_{i\tau }$ for $%
\tau >t$, are uncorrelated with $x_{it}$. \

Some work in this literature has been willing to make such timing (and
information set) assumptions stronger. \ For example, in OP/LP/ACF
procedures, it is often assumed that firms need to commit to their capital
choice at $t-1$. \ In demand models considering endogenous product
characteristics, it is sometimes assumed that those product design decisions
are made one or more periods before the product is sold. \ If one assumes
that $x_{it}$ is chosen at $t-1$ (and one continues to make the assumption
that firms do not observe $\omega_{it}$ and $\kappa_{it}$ until $t$) then
information set $I_{it-1}$ in Assumption 1 additionally includes $x_{it}$. \
This means that (\ref{eq:theequation}) implies an additional orthogonality
condition that can be used for estimation (since $\xi_{it}$ is now mean
independent of $x_{it}$). Intuitively, since $x_{it}$ is chosen at $t-1$,
innovations that are not observed until after that choice, e.g. $\xi _{it}$
are orthogonal to it.

In the dynamic panel literature, this assumption is often referred to as $%
x_{it}$ being ``predetermined''. \ Ackerberg (2020) details how these timing
and information set assumptions can be strengthened or relaxed -- an
important point is that it is not only the timing of when $x_{it}$ is chosen
that matters but also what the firm knows at that time. \ For example, if $%
x_{it}$ is chosen at $t-1$ but one assumes that firms observe $\omega _{it}$
and $\kappa _{it}$ also in $t-1$ (one period earlier), then one does not
obtain additional moment restrictions.

So, consider an alternative model where firms make input decisions at time $%
t-1$ as a function of their information set $I_{i,t-1}$, e.g.

\begin{equation}
x_{it}=\pi ^{0}+\theta ^{0}\rho _{\omega }^{0}\omega _{i,t-1}+\kappa
_{i,t-1}.  \label{fixedinput}
\end{equation}%
Note that $\rho _{\omega }^{0}\omega _{i,t-1}$ is a conditional expectation
of $\omega _{it}$ at time $t-1$. \ In a production function context, it
makes sense that a firm would base their $x_{it}$ decision on their
expectation of $\omega _{it}$, but this assumption is not necessary as our
derivation is for arbitrary $\theta ^{0}$. Since $x_{it}$ is chosen at $t-1,$
it is now a function of $\kappa _{i,t-1}$ instead of $\kappa _{it}$
(because, like $\omega _{it}$, $\kappa _{it}$ has not been observed yet).%
\footnote{%
Working with $\kappa _{it}$ instead of $\kappa _{i,t-1}$ does not change our
result (i.e. firms knowing $\kappa _{it}$ at time $t-1$). What matters for
our result is firm $i$ knowing only $\omega _{i,t-1}$ at time $t-1$, not $%
\omega _{it}$. It does not matter for $\kappa _{it}$. In some sense it is
just a relabeling of $\kappa _{it}$. More generally, what is important for a
stronger timing/information assumption to rule out the pseudo-solution is
the unobserved factor that enters both the productivity function and the
input function, not for the unobserved factor(s) that only enters the input
function.}

Following the arguments of the benchmark case, we can consider the basic
dynamic panel moment evaluated at the pseudo-parameters $(\beta,\rho
_{\omega },\alpha )=(\beta ^{0}+\frac{1}{\theta ^{0}},\rho _{x}^{0},\alpha
^{0}-\frac{\pi ^{0}}{\theta ^{0}})$. \ The first part of the derivation from
Section 2 stays the same:

\begin{align}
& E[(y_{it}-\rho _{\omega }y_{i,t-1})-\alpha (1-\rho _{\omega })-\beta
(x_{it}-\rho _{\omega }x_{i,t-1})|I_{i,t-1}]  \label{newder} \\
& =E[(y_{it}-\rho _{x}^{0}y_{i,t-1})-\left( \alpha ^{0}-\frac{\pi ^{0}}{%
\theta ^{0}}\right) (1-\rho _{x}^{0})-\left( \beta ^{0}+\frac{1}{\theta ^{0}}%
\right) (x_{it}-\rho _{x}^{0}x_{i,t-1})|I_{i,t-1}]  \notag \\
& =E[\alpha ^{0}(1-\rho _{x}^{0})+\beta ^{0}(x_{it}-\rho
_{x}^{0}x_{i,t-1})+\omega _{it}-\rho _{x}^{0}\omega _{i,t-1}+\eta _{it}-\rho
_{x}^{0}\eta _{i,t-1}  \notag \\
& -\left( \alpha ^{0}-\frac{\pi ^{0}}{\theta ^{0}}\right) (1-\rho
_{x}^{0})-\left( \beta ^{0}+\frac{1}{\theta ^{0}}\right) (x_{it}-\rho
_{x}^{0}x_{i,t-1})|I_{i,t-1}]  \notag \\
& =E[\frac{\pi ^{0}}{\theta ^{0}}(1-\rho _{x}^{0})-\frac{1}{\theta ^{0}}%
(x_{it}-\rho _{x}^{0}x_{i,t-1})+\omega _{it}-\rho _{x}^{0}\omega
_{i,t-1}+\eta _{it}-\rho _{x}^{0}\eta _{i,t-1}|I_{i,t-1}].  \notag
\end{align}

However, $(x_{it}-\rho _{x}^{0}x_{i,t-1})$ is now different. \ Using (\ref%
{fixedinput}), we have
\begin{equation*}
(x_{it}-\rho _{x}^{0}x_{i,t-1})=\left( 1-\rho _{x}^{0}\right) \pi
^{0}+\theta ^{0}\left[ \rho _{\omega }^{0}\omega _{i,t-1}-\rho _{x}^{0}\rho
_{\omega }^{0}\omega _{i,t-2}\right] +\left( \kappa _{i,t-1}-\rho
_{x}^{0}\kappa _{i,t-2}\right).
\end{equation*}%
and note also that we can write%
\begin{equation*}
\omega _{it}-\rho _{x}^{0}\omega _{i,t-1}=\rho _{\omega }^{0}\omega
_{i,t-1}-\rho _{x}^{0}\rho _{\omega }^{0}\omega _{i,t-2}+\xi _{it}-\rho
_{x}^{0}\xi _{i,t-1}.
\end{equation*}%
Substituting these into (\ref{newder}), we get
\begin{eqnarray*}
&&E[\frac{\pi ^{0}}{\theta ^{0}}(1-\rho _{x}^{0})-\frac{1}{\theta ^{0}}%
(x_{it}-\rho _{x}^{0}x_{i,t-1})+\omega _{it}-\rho _{x}^{0}\omega
_{i,t-1}+\eta _{it}-\rho _{x}^{0}\eta _{i,t-1}|I_{i,t-1}] \\
&=&E[\frac{\pi ^{0}}{\theta ^{0}}(1-\rho _{x}^{0})-\frac{1}{\theta ^{0}}%
\left[ \left( 1-\rho _{x}^{0}\right) \pi ^{0}+\theta ^{0}\left[ \rho
_{\omega }^{0}\omega _{i,t-1}-\rho _{x}^{0}\rho _{\omega }^{0}\omega _{i,t-2}%
\right] +\left( \kappa _{i,t-1}-\rho _{x}^{0}\kappa _{i,t-2}\right) \right]
\\
&&+\rho _{\omega }^{0}\omega _{i,t-1}-\rho _{x}^{0}\rho _{\omega }^{0}\omega
_{i,t-2}+\xi _{it}-\rho _{x}^{0}\xi _{i,t-1}+\eta _{it}-\rho _{x}^{0}\eta
_{i,t-1}|I_{i,t-1}] \\
&=&E[-\frac{1}{\theta ^{0}}u_{i,t-1}+\xi _{it}-\rho _{x}^{0}\xi
_{i,t-1}+\eta _{it}-\rho _{x}^{0}\eta _{i,t-1}|I_{i,t-1}].
\end{eqnarray*}

Importantly, because $x_{it}$ is chosen at $t-1$, it is in $I_{i,t-1}$. \
And because $x_{it}$ is chosen as a function of $\omega _{i,t-1}$ and $%
\kappa _{i,t-1}$, it is clearly correlated with $u_{i,t-1}$ and $\xi _{it-1}$%
. \ Hence, as long as $x_{it}$ is utilized in the formation of moment
conditions, the pseudo-solution can be ruled out.\footnote{
This does not necessarily preclude possible pseudo-solutions at other points
in the parameter space, but we have not found any in our simulations (nor
have we found local minimums).} Of course, at the true parameters, the
dynamic panel moment condition is
\begin{eqnarray}
&&E[(y_{it}-\rho _{\omega }^{0}y_{i,t-1})-\alpha ^{0}(1-\rho _{\omega
}^{0})-\beta ^{0}(x_{it}-\rho _{\omega }^{0}x_{i,t-1})|I_{i,t-1}]  \notag \\
&=&E[(y_{it}-\alpha ^{0}-\beta ^{0}x_{it})-\rho _{\omega
}^{0}(y_{i,t-1}-\alpha ^{0}-\beta ^{0}x_{i,t-1})|I_{i,t-1}]  \notag \\
&=&E[\omega _{it}-\rho _{\omega }^{0}\omega _{i,t-1}+\eta _{it}-\rho
_{\omega }^{0}\eta _{i,t-1}|I_{i,t-1}]=E[\xi _{it}+\eta _{it}-\rho _{\omega
}^{0}\eta _{i,t-1}|I_{i,t-1}]=0,  \notag
\end{eqnarray}%
and since $\xi _{it}$ is not in $I_{i,t-1}$ and hence not correlated with $%
x_{it}$, this continues to hold. \

In conclusion, if one is willing to make stronger timing assumptions, it
appears that the pseudo-solution problem we have identified disappears. \
Presumably this is also the case if one is willing to make even stronger
timing assumptions, e.g. where $x_{it}$ is chosen at some $t-\Delta $ where $%
\Delta $ $>1$. \ On the other hand, and as evidenced by the empirical
literature, these are often assumptions that researchers are unwilling to
make, such as in the case of variable inputs in the production function
context.


\section{Identification}
\label{Sec4}
We now turn to strategies to resolve this under-identification
problem. First, we focus on the benchmark model from Section 2 (i.e. linear
$x_{it}$ process with AR(1) shock) and derive a system of reduced form
equations that provides an alternative representation of the
under-identification problem. Second, we show how, under a sign restriction
on the structural parameter of the input process, we can use this reduced
form representation to form an identification strategy. Estimation follows
straightforwardly. Third, we use the intuition from this simple model to
propose several practical strategies in more general models.

\subsection{Reduced form representation}
\label{Sec41}
We now re-visit the baseline full model in Section 2
(equations (\ref{eq:pf0}),(\ref{eq:x_w}),(\ref{eq:inputeq}) and (\ref{eq:x_ar})).
These equations imply the following system of ``reduced-form''
equations for $y_{it}$ and $x_{it}$, where the explanatory variables are
dated at period t-1 or before\footnote{We thank Victor Aguirregabiria for suggesting this alternative
representation.}
\begin{eqnarray}
y_{it} &=&\pi _{y0}+\pi _{yy}y_{i,t-1}+\pi _{yx}x_{i,t-1}+e_{it}^{y},
\label{redformy} \\
x_{it} &=&\pi _{x0}+\pi _{xy}y_{i,t-1}+\pi _{xx}x_{i,t-1}+e_{it}^{x},
\label{redformx}
\end{eqnarray}%
and where $e_{it}^{y}=\beta (\theta \xi _{it}+u_{it})+\xi _{it}+\eta
_{it}-\rho _{\omega }\eta _{i,t-1}-\beta \theta (\rho _{\omega }-\rho
_{x})\eta _{i,t-1}$ and $e_{it}^{x}=\theta \xi _{it}+u_{it}-\theta (\rho
_{\omega }-\rho _{x})\eta _{i,t-1}$. The six reduced-form parameters in
(\ref{redformy})-(\ref{redformx}) relate to the six
structural parameters $(\beta,\theta,\rho_{\omega},\rho _{x},\alpha,\pi
)$ according to the equation system \eqref{sys:pi} in the Appendix.

It is straightforward to consistently estimate the reduced-form parameters
in equations (\ref{redformy}) and (\ref{redformx}). According to our
assumptions, the two error terms are mean independent of $x_{i,t-1}$, but
not of $y_{i,t-1}$ (since $y_{i,t-1}$ is a function of $\eta _{i,t-1}$).
However, under our assumptions, $y_{i,t-2}$ is a valid instrument for $%
y_{i,t-1}$ and can be used for IV estimation. Given IV estimates of the
reduced form parameters $\pi $, we can recover the structural parameters $%
(\beta ,\theta ,\rho _{\omega },\rho _{x},\alpha ,\pi )$ by inverting the
system of parameter equations (\eqref{sys:pi}). However, analogous to our
derivations in Section 2, there are two solutions of the mapping from the
reduced form parameters to the structural parameters. Notably, the values of
$\theta $ at the two solutions are:
\begin{eqnarray}
\theta  &=&+\frac{\pi _{xy}}{\sqrt{(\pi _{yy}-\pi _{xx})^{2}+4\pi _{yx}\pi
_{xy}}}  \label{twothetas} \\
\theta  &=&-\frac{\pi _{xy}}{\sqrt{(\pi _{yy}-\pi _{xx})^{2}+4\pi _{yx}\pi
_{xy}}}  \notag
\end{eqnarray}%
and the Appendix shows, for each of these possible $\theta$'s, there is a solution
of the mapping from the reduced form parameters into
the remaining structural parameters. The fact that (\ref{twothetas}) imply that the $\theta$'s at the two
possible solutions are guaranteed to have opposite sign will be useful to us
momentarily.\footnote{This reduced form approach
generalizes to the case of multiple endogenous variables. In this case,
the mapping from the reduced form to the structural parameters is again from
$R^{d}$ to $R^{d}$ (where d is the number of structural parameters) but now
contains multiple quadratic equations, and, as illustrated in Section 3.2,
there is typically more than one spurious solution.}

This reduced form representation is an alternative way of expressing the
under-identification issue studied in this paper - one can easily verify
that for the parameters $(\beta,\rho_{\omega},\alpha )$, the relationship
between the two solutions\ are exactly the same as that described by Theorem
1. \ However, we think that this reduced form setup is theoretically
interesting in itself because it shows that the under-identification issue
can be alternatively represented in terms of a mapping from reduced form to
structural parameters (rather than multiple solutions to a set of sample
moments). \ Notably, this result means that for essentially \emph{any}
dataset, if one estimates this reduced form model (derived from the
structural model (\ref{eq:pf0}),(\ref{eq:x_w}),(\ref{eq:inputeq}) and (\ref{eq:x_ar})),
there will \textit{necessarily} be two solutions. \ In other
words, there will be two solutions \textit{regardless} of the true DGP
underlying the data. \ This seems counter to some of the results in Section
3, where we showed that for some DGPs (e.g. with stronger timing
assumptions), the standard dynamic panel moments appear to not have multiple
solutions. \ This is because there is a difference between the exact moments
used in Sections 2 and 3 versus those used in the reduced form approach -
those in the reduced form approach necessarily impose the input choice
equation for all the moment conditions. \ This result also points to the
pervasiveness of the potential under-identification problem in these types
of models - one can get under-identification either due to the specific
moments used (as in this section), or due to the DGP (as illustrated in
Sections 2 and 3 using more traditional moments).\ While the latter might be
most important in practice, the former also seems important to be aware of
since researchers often experiment with different sets of moments in these
types of problems.

This reduced form representation is also useful because it provides
estimates of the parameters of the process determining $x_{it}$, i.e. $%
(\theta,\rho_{x},\pi)$ and we next show how this, plus a potential sign
restrictions on one of these parameters, provides one potential solution to
the under-identification problem.

\subsection{Sign Restriction}

\label{Sec42}

The above derivation indicates that the two solutions for $\theta $ produced
by the reduced form approach necessarily have the opposite sign. \ Recall
that $\theta ^{0}$ measures how the endogenous $x_{it}$\ depends on the
unobserved term $\omega _{it}$, i.e. $x_{it}=\pi ^{0}+\theta ^{0}\omega
_{it}+\kappa _{it}$. Thus, its sign represents the direction of the
endogeneity problem. In many cases one may be willing to \textit{a priori}
sign $\theta ^{0}$ (e.g., in the production function context, firm profit
maximization would likely imply $\theta ^{0}>0$; in other econometric
contexts, one might be willing to sign the endogeneity problem (e.g., Manski
and Pepper (2000)). In these cases, the reduced form representation provides
a simple two-step method to consistently estimate the structural parameters,
i.e. estimate (\ref{redformy}) and (\ref{redformx}) using IV estimation, use the
mapping between the reduced form and the structural parameters to calculate
the two sets of structural parameters, and choose the appropriate one given
the sign restriction on $\theta ^{0}$.

In fact, a sign restriction can help resolve the under-identification
problem even when we only consider moment condition (\ref{eq:theequation}),
as more commonly done in the traditional dynamic panel approach. In Section
2, we showed that both $A=(\alpha ^{0},\beta ^{0},\rho _{\omega }^{0})$ and $%
B=(\alpha ^{0}-\frac{\pi ^{0}}{\theta ^{0}},\beta ^{0}+\frac{1}{\theta ^{0}}%
,\rho _{x}^{0})$ solve the conditional moment condition %
\eqref{eq:theequation} given the information set $I_{it}$. Since
element-wise differencing $A-B$ gives $(-\frac{\pi ^{0}}{\theta^{0}},-\frac{%
1}{\theta ^{0}},\rho _{\omega }^{0}-\rho _{x}^{0})$, the difference between
the two solutions for $\beta $ depends only on $\theta^{0}$ - the
structural parameter in the input process. Therefore, if we find both
solutions to the moment condition, knowledge of the sign of $\theta^{0}$
tells us which of the two solutions is the correct one. In other words, if we
assume $\theta^{0}>0$, then the coefficient vector with the lower estimate
of $\beta ^{0}$ is the true solution (and vice-versa).\footnote{
In some cases, it may be more appealing to make sign restrictions on other
parameters, e.g. on $\beta $. Unlike $\theta $, there is no guarantee that
the true and pseudo-solutions for $\beta $ have opposite sign. However,
they can have opposite signs (if $\beta ^{0}$ and $\theta ^{0}$ have
opposite signs and $\left\vert \frac{1}{\theta ^{0}}\right\vert >\left\vert
\beta^{0}\right\vert)$, and this could provide the basis for identifying
the pseudo-solution. In any case, we should be aware that the key to
utilizing sign restrictions is finding both minima first.}

\subsection{More General Lessons}

\label{sec:lesson}

Sections \ref{Sec41} and \ref{Sec42} suggest a reduced form approach
together with a sign restriction to solve the under-identification problem
in our simple model. However, this relies on the specific DGP for $x_{it}$
given by (\ref{eq:inputeq}) and (\ref{eq:x_ar}). Given that the previous
section suggests that these dynamic panel and proxy variable approaches have
similar identification issues with more general DGPs for $x_{it}$, a natural
question is what to do in those cases. In particular, suppose a researcher
is \textit{not} willing to impose (\ref{eq:inputeq}) and (\ref{eq:x_ar}) and
instead wants to allow for (or model) a more general $x_{it}$ process. \
Obviously, it is difficult for us to give exact advice given the many
possible models that researchers may construct. However, we feel our
solution in the simple case provides some constructive advice on how
researchers might reduce the likelihood of ending up at a pseudo-solution in
more complex models.

One possibility would be to specify the DGP for $x_{it}$ up to parameters
and follow the logic of Sections \ref{Sec41} and \ref{Sec42} to derive an
analogous system of estimating equations. Given the intuition above,
imposing assumptions on the direction of the relationship between $x_{it}$
and $\omega _{it}$, analogous to the sign restriction on $\theta $ used
above, could resolve the under-identification problem.

However, suppose one is unwilling to commit to assuming a specific model for
$x_{it}$. This is perhaps more natural given the history of this literature -
one nice aspect of both the dynamic panel and proxy variable methods is that
they do \textit{not} rely on fully specifying a parametric DGP for $x_{it}$
(though often they do make some assumptions on these DGPs, e.g. the scalar
unobservable assumption used in the proxy variable literature). \ We also
believe our findings lead to constructive practical advice in this
situation. In particular, in our simple model (and in ``nearby'' more complex models),
the pseudo-solution for $\beta$ is greater than the true $\beta^{0}$ (this is based on $\theta
^{0}>0$, the reverse is true when $\theta^{0}<0$). This means that we
know the direction in which we might err. In particular, if one has found
a candidate estimate $\widehat{\beta}$ (and believes $\theta^{0}>0$), one
might want to be extra diligent in continuing to search for minima at
parameters where $\beta <\widehat{\beta}$. This is because if one has
erroneously found the spurious zero, the non-pseudo-solution for $\beta$
should be below $\widehat{\beta}$. One should of course always be
diligent in numeric search to avoid local minima, but the results here
suggest one should be extra concerned about local minima in one specific
direction relative to a candidate estimate $\widehat{\beta}$.

Another possibility is to estimate the model based on only the traditional
dynamic panel moment condition (\ref{eq:theequation}), i.e. without
specifying a DGP for $x_{it}$, but to then do some ex-post testing of that
solution. In particular, given a candidate estimate $(\widehat{\alpha},
\widehat{\beta},\widehat{\rho}_{\omega})$ based on moment condition (\ref
{eq:theequation}), one can construct the residuals
\begin{equation}
\widehat{\omega_{it}+\eta _{it}}=y_{it}-\widehat{\alpha }-\widehat{\beta }%
x_{it}.\notag
\end{equation}%
Again, assuming one is willing to sign $\theta$, one could examine the
relationship between $x_{it}$ and this residual, e.g. through a regression
or correlation analysis. Although this regression cannot be interpreted
causally because of the existence of $\eta_{it}$ in the constructed
residual, the measurement error tends to attenuate the coefficient.\footnote{
Note that this would be easier in the proxy variable methods, as a
by-product of the assumptions guaranteeing invertibility in those models
allows one to separately identify $\widehat{\omega _{it}}$ and $\widehat{
\eta_{it}}$.} \ So, at least in models ``close'' to our simple model, one would expect that the sign of
this relationship at the pseudo-solution would be the reverse of what one
would expect. \ In other words, if one believes that $\theta>0$, an
indication of a possible pseudo-solution would be finding a negative
correlation between $\widehat{\omega _{it}+\eta _{it}}$ and $x_{it}$.\footnote{
Note that this is related to the practice of comparing these coefficient
estimates to those of standard OLS, i.e. assessing signing the bias of OLS.
This is because the sign of the OLS bias is related to $\theta $. So the
possibility of pseudo-solutions provides another rationale for doing this,
i.e. if one is at the pseudo-solution, the sign of the OLS bias may be the
opposite of what one expects it to be.} Alternatively, one can impose the
sign restriction as a moment inequality: $E[x_{it}(y_{it}-\alpha ^{0}-\beta
^{0}x_{it})]\geq 0$.

Lastly, one might consider starting by estimating a model that is more
robust to the pseudo-solution problem but is not necessarily an appropriate
model. \ Initial estimates from this misspecified model can be used as
starting values for estimating the correct model to reduce the likelihood of
ending up at a pseudo-solution in the correct model. \ Our above analyses
suggest at least two possibilities for this. \ First, even if one does not
believe the stronger timing assumption of Section \ref{sec:stronger}, one
might estimate that model anyway. Then one could use the estimates as
starting values for a model with the (correct) weaker timing assumption. \
The hope is that the stronger timing assumption does not overly bias the
parameters, and thus keeps the numeric search procedure from straying to a
pseudo-solution. \ \ Second, one could assume input choice (\ref{eq:inputeq}%
) and use our reduced-form procedure and the sign restriction to recover
structural parameters, even if one does not believe that the input choice
equation is exactly specified. \ One could again use the resulting estimates
as starting parameters for an estimation procedure that does not impose a
specific input choice equation. \ Again, the hope is that while those
starting parameters may be inconsistent, they may be close enough to the
true minima to be helpful at getting the numeric search procedure to get the
true solutions as long as the input process is not too far away from the
linear input demand model.

Of course, these informal suggestions are not foolproof in avoiding
pseudo-solutions, and in some datasets with even more complex $x_{it}$
processes, there may not even be pseudo-solutions. \ However, we feel that
being aware of the possibility of such pseudo-solutions and this guidance
for avoiding them can be useful for applied researchers.

\section{Conclusion}

We have studied a potential identification problem in methods that use
timing and information set assumptions to resolve endogeneity problems in
structural models. These methods have been applied widely, both in the
production function context and elsewhere. Prior work calls attention to
this identification problem in a fairly narrow context. We show that the
identification problem appears to be endemic across a broad set of models
using these types of assumptions, and demonstrate how to resolve it using a
sign restriction and additional modelling. We then suggest
ways of extracting intuition from that solution to more complex empirical
settings.


\renewcommand{\theequation}{A.\arabic{equation}}

\setcounter{equation}{0}

\section*{Appendix}

\begin{proof}[Proof of Theorem 1]
To see the pseudo-solution in the benchmark model, consider moment condition (
\ref{eq:theequation}), where $\rho_{\omega}$ is incorrectly set at $\rho
_{x}^{0}$, $\alpha $ is incorrectly set at $\alpha ^{0}-\frac{\pi ^{0}}{
\theta^{0}}$, and $\beta$ is incorrectly set at $\beta ^{0}+\frac{1}{
\theta^{0}}:$
\begin{align*}
& E[(y_{it}-\rho _{\omega }y_{i,t-1})-\alpha (1-\rho _{\omega })-\beta
(x_{it}-\rho _{\omega }x_{i,t-1})|I_{i,t-1}]    \numberthis \label{eq:theproof} \\
& =E[(y_{it}-\rho _{x}^{0}y_{i,t-1})-\left( \alpha ^{0}-\frac{\pi ^{0}}{%
\theta ^{0}}\right) (1-\rho _{x}^{0})-\left( \beta ^{0}+\frac{1}{\theta ^{0}}%
\right) (x_{it}-\rho _{x}^{0}x_{i,t-1})|I_{i,t-1}]  \notag \\
& =E[\alpha ^{0}(1-\rho _{x}^{0})+\beta ^{0}(x_{it}-\rho
_{x}^{0}x_{i,t-1})+\omega _{it}-\rho _{x}^{0}\omega _{i,t-1}+\eta _{it}-\rho
_{x}^{0}\eta _{i,t-1}  \notag \\
& -\left( \alpha ^{0}-\frac{\pi ^{0}}{\theta ^{0}}\right) (1-\rho
_{x}^{0})-\left( \beta ^{0}+\frac{1}{\theta ^{0}}\right) (x_{it}-\rho
_{x}^{0}x_{i,t-1})|I_{i,t-1}]  \notag \\
& =E[\frac{\pi ^{0}}{\theta ^{0}}(1-\rho _{x}^{0})-\frac{1}{\theta ^{0}}%
(x_{it}-\rho _{x}^{0}x_{i,t-1})+\omega _{it}-\rho _{x}^{0}\omega
_{i,t-1}+\eta _{it}-\rho _{x}^{0}\eta _{i,t-1}|I_{i,t-1}]  \notag \\
& =E[\frac{\pi ^{0}}{\theta ^{0}}(1-\rho _{x}^{0})-\frac{1}{\theta ^{0}}%
(\left( 1-\rho _{x}^{0}\right) \pi ^{0}+\theta ^{0}\left( \omega _{it}-\rho
_{x}^{0}\omega _{i,t-1}\right) +\left( \kappa _{it}-\rho _{x}^{0}\kappa
_{i,t-1}\right) )  \notag \\
& +\omega _{it}-\rho _{x}^{0}\omega _{i,t-1}+\eta _{it}-\rho _{x}^{0}\eta
_{i,t-1}|I_{i,t-1}]  \notag \\
& =E[-\frac{1}{\theta ^{0}}\left( \kappa _{it}-\rho _{x}^{0}\kappa
_{i,t-1}\right) +\eta _{it}-\rho _{x}^{0}\eta _{i,t-1}|I_{i,t-1}]  \notag \\
& =E[-\frac{1}{\theta ^{0}}u_{it}+\eta _{it}-\rho _{x}^{0}\eta
_{i,t-1}|I_{i,t-1}]=0,  \notag
\end{align*}%
where the second and fourth equalities come from substituting in the true
DGP from (\ref{eq:pf0}) and (\ref{eq:inputeq}) respectively.
\end{proof}

\begin{proof}[Proof of Equations \eqref{redformy} and \eqref{redformx}]
Consider the baseline full model below. We derive a system of reduced form equations of $x_t$ and $y_t$. We omit firm index $i$ for ease of notation.
\begin{equation}
\begin{aligned}\label{y}
y_{t} &=\alpha^0 + \beta^0 x_{t}+\omega_{t}+\eta _{t}, \\
x_{t} &=\pi^0 + \theta^0\omega^0_{t}+\kappa_{t},  \\
\omega_{t} &=\rho^0_{\omega}\omega _{t-1}+\xi _{t}, \\
\kappa _{t} &=\rho^0_{x}\kappa _{t-1}+u_{t}.
\end{aligned}
\end{equation}

We first obtain the reduced form equation for $x_t$
\begin{eqnarray*}
x_{t} &=&\pi^0 + \theta^0\omega _{t}+\kappa _{t} \\
&=&\pi^0 + \theta^0(\rho^0_{\omega}\omega_{t-1}+\xi _{t}) +\rho^0_{x
}\kappa _{t-1}+u_{t} \\
&=&\pi^0 + \theta^0( \rho^0_{\omega}\omega _{t-1}+\xi _{t}) +\rho^0_{x}( x_{t-1}-\pi^0-\theta^0\omega _{t-1}) +u_{t} \\
&=&(1-\rho^0_{x})\pi^0 + \theta^0( \rho^0_{\omega}-\rho^0_{x}) \omega _{t-1}+\rho^0_{x}x_{t-1}+\theta^0\xi_{t}+u_{t} \\
&=&(1-\rho^0_{x})\pi^0+ \theta^0(\rho^0_{\omega}-\rho^0_{x})(y_{t-1}-\alpha^0 - \beta^0
x_{t-1}-\eta_{t-1}) + \rho^0_{x}x_{t-1}+\theta^0\xi_{t}+u_{t}\\
&=&\pi_{x0} + \theta^0(\rho^0_{\omega}-\rho^0_{x})y_{t-1}+(\rho^0_{x
}-\beta^0\theta^0(\rho^0_{\omega }-\rho^0_{x}))x_{t-1} + e_{t}^{x},
\end{eqnarray*}%
where $\pi_{x0} = (1-\rho^0_{x})\pi^0 - \theta^0(\rho^0_{\omega}-\rho^0_{x})\alpha^0$ and $e_{t}^{x} = \theta^0\xi_{t}+u_{t}-\theta^0(\rho^0_{\omega}-\rho^0_{x})\eta_{t-1}$.

$\rho^0_{\omega}$-difference the production function and plug in the reduced-form equation of $x_t$, resulting in the reduced form equation for $y_t$
\begin{align*}
y_{t} &=\alpha^0(1-\rho^0_{\omega}) + \rho^0_{\omega}y_{t-1}+\beta^0x_{t}-\rho^0_{\omega}\beta^0x_{t-1}+\xi
_{t}+\eta_{t}-\rho^0_{\omega }\eta_{t-1} \\
&=\alpha^0(1-\rho^0_{\omega}) + \rho^0_{\omega}y_{t-1}+\beta^0[\theta^0(\rho^0_{\omega}-\rho^0
_{x})y_{t-1}+(\rho^0_{x}-\beta^0\theta^0(\rho^0_{\omega
}-\rho^0_{x}) )x_{t-1}\\
& + \pi_{x0} + e_{t}^{x}] -\rho^0_{\omega}\beta^0
x_{t-1}+\xi_{t}+\eta_{t}-\rho^0_{\omega}\eta _{t-1} \\
&= \pi_{y0} + [\rho^0_{\omega}+\beta^0\theta^0(\rho^0_{\omega}-\rho^0_{x})] y_{t-1}+\beta^0[(1+\beta^0\theta^0)(\rho^0_{x}-\rho^0_{\omega})] x_{t-1}+e_{t}^{y},
\end{align*}
where $\pi_{y0} = \beta^0(1 - \rho^0_{x})\pi^0 + [(1-\rho^0_{\omega}) - \theta^0\beta^0(\rho^0_{\omega}-\rho^0_{x}))]\alpha^0$
and $e_{t}^{y} = \beta^0(\theta^0 \xi _{t}+u_{t}) +\xi_{t}+\eta _{t}-\rho^0_{\omega}\eta _{t-1}-\beta^0
\theta^0(\rho^0_{\omega}-\rho^0_{x})\eta _{t-1}$.

In summary, we have
\begin{eqnarray}
y_{t} &=&\pi_{y0} + \pi _{yy}y_{t-1}+\pi _{yx}x_{t-1}+ \beta(\theta \xi _{t}+u_{t}) +\xi _{t}+\eta _{t}-\rho _{\omega }\eta _{t-1}-\beta
\theta(\rho _{\omega}-\rho _{x}) \eta _{t-1},\notag \\
x_{t} &=&\pi_{x0} + \pi _{xy}y_{t-1}+\pi_{xx}x_{t-1}+ \theta \xi
_{t}+u_{t}-\theta(\rho _{\omega }-\rho _{x}) \eta _{t-1}, \notag
\end{eqnarray}%
where
\begin{equation}
\begin{aligned}\label{sys:pi}
\pi _{yy} &=\rho_{\omega}+\beta \theta(\rho _{\omega }-\rho
_{x}),  \\
\pi _{yx} &=-\beta( 1+\beta \theta)(\rho _{\omega
}-\rho _{x}),  \\
\pi _{xy} &=\theta(\rho_{\omega}-\rho _{x}), \\
\pi _{xx} &=\rho _{x}-\beta \theta(\rho_{\omega }-\rho
_{x}), \\
\pi_{y0} &= \beta(1 - \rho_{x})\pi + [(1-\rho_{\omega}) - \theta\beta(\rho_{\omega}-\rho_{x}))]\alpha,\\
\pi_{x0} &= (1-\rho_{x})\pi - \theta (\rho_{\omega}-\rho_{x})\alpha .
\end{aligned}%
\end{equation}

Inverting the system of equations gives the corresponding solutions for the remaining parameters once $\theta$ is identified
\begin{align*}
\beta &= \frac{1}{2}(\frac{\pi_{yy}-\pi_{xx}}{\pi_{xy}} - \frac{1}{\theta}),\\
\rho_{\omega} & = \frac{1}{2}(\pi_{yy}+\pi_{xx} + \frac{\pi_{xy}}{\theta}),\\
\rho_{x} &= \frac{1}{2}(\pi_{yy}+\pi_{xx} - \frac{\pi_{xy}}{\theta}),\\
\pi & = \frac{\pi_{x0}\pi_{xy}+(1-\pi_{yy})\pi_{y0}}{1-(\pi_{xx}+\pi_{yy})+(\pi_{xx}\pi_{yy}-\pi_{yx}\pi_{xy})},\\
\alpha & = \frac{(1-\rho_{x})\pi - \pi_{y0}}{\pi_{xy}}.
\end{align*}

\end{proof}

\end{document}